\documentclass[onecolumn,floatfix,superscriptaddress,showpacs,showkeys,nofootinbib,preprint]{revtex4} 
\usepackage{epsfig}
\usepackage{amssymb,latexsym,amsmath}

\begin{document}
\title{Multiplicity Fluctuations in Limited Segments of Momentum Space\\ in
  Statistical Models }
\author{Michael Hauer}
\affiliation{Helmholtz Research School, University of Frankfurt, Frankfurt,
  Germany} 

\begin{abstract}
Multiplicity fluctuations in limited segments of momentum space are calculated
for a classical pion gas within the statistical model. Results for the grand
canonical, canonical, and micro-canonical ensemble are obtained, compared and
discussed. We demonstrate that even in the large volume limit correlations
between macroscopic subsystems due to energy and momentum conservation
persist. Based on the micro-canonical formulation we make qualitative
predictions for the rapidity and transverse momentum dependence of
multiplicity fluctuations. The resulting effects are of similar magnitude as
the predicted enhancement due to a phase transition from a quark-gluon plasma
to a hadron gas phase, or due to the critical point of strongly interacting
matter, and qualitatively agree with recently published preliminary
multiplicity fluctuation data of the NA49 SPS experiment.    
\end{abstract}



\maketitle
\section{Introduction}
\label{Intro}
The statistical model has been, for a long time, successfully  applied to 
fit experimental data on mean hadron multiplicities in heavy ion collision
experiments over a wide range of beam energies and system sizes. For recent
reviews see \cite{FOCley,FOBeca,FOPBM,FORafe}. So naturally the question arises
whether the statistical model is able to describe event-by-event fluctuations
of these observables as well. And indeed, a first comparison suggests that
this might be possible for the sample of most central events. Global
conservation laws, imposed on a statistical system, lead, even in  the large
volume limit, to suppressed fluctuations. The multiplicity distributions of
charged hadrons recently reported \cite{NA49_fluc} by the NA49 SPS experiment
are systematically narrower than a Poissonian reference distribution. This
could be interpreted \cite{MCEvsData} as effects due to energy and charge
conservation in a relativistic hadronic gas.   

Multiplicity fluctuations are usually quantified by the ratio of the variance
of a 
multiplicity distribution to its mean value, the so-called scaled variance. In
statistical models there is a qualitative difference in the
properties of mean value and scaled variance. In the case of the mean
multiplicity results obtained within the grand canonical ensemble (GCE),
canonical ensemble (CE), and micro-canonical ensemble (MCE)  approach  each
other in the large volume limit. One refers here to as the thermodynamic
equivalence of these ensembles. It was recently found \cite{CEfluc1} that
corresponding results for the scaled variance are different in different
ensembles, and thus this observable is sensitive to conservation laws obeyed
by a statistical system. 

The growing interest in the experimental and theoretical study of fluctuations
in strong interactions (see e.g., reviews \cite{reviewfluc}) is motivated by
expectations of anomalies in the vicinity of the onset of deconfinement
\cite{ood} and in the case when the expanding system goes through the
transition line between a quark-gluon plasma and a hadron gas phase
\cite{phasetrans}. In particular, a critical point of strongly interacting
matter may be accompanied by a characteristic power-law pattern in fluctuations
\cite{critpoint}. A non-monotonic dependence of event-by-event fluctuations on
system size and/or center of mass energy in heavy ion collisions would
therefore give valuable insight into the phase diagram of strongly
interacting matter. Provided the signal survives the subsequent evolution and
hadronization of the system (see also \cite{recomb}). Therefore, in order to
asses the discriminating power of proposed measures, for a recent review see
\cite{reviewfluc2}, one should firstly  study properties of equilibrated
sources \cite{MCEvsData,res,VolDep,vdw} and quantify `baseline` (or
thermal/statistical) fluctuations. Apart from being an important tool in an
effort to study a possible critical behavior, the study of fluctuations within
the statistical model constitutes also a further test of its validity.  

In this paper we make detailed predictions for the momentum space
dependence of multiplicity fluctuations. We show that energy and momentum
conservation lead to a non-trivial dependence of the scaled variance on the
location and magnitude of the observed fraction of momentum space. 
These predictions can be tested against existing and future data from the
heavy ion collision experiments at the CERN SPS and BNL RHIC facilities.   

The paper is organized as follows: In section \ref{model} we briefly
introduce our model. In section \ref{GCECE} we consider multiplicity
distributions in a limited region of momentum space in GCE and CE. For the MCE
we follow, in section \ref{MCE},  the procedure of Ref.\cite{clt} and
show how to calculate the width of the corresponding distributions in the
large volume limit. We revisit the so-called `acceptance scaling` previously
suggested as an approximate implementation of experimental acceptance in
section \ref{Results}. Technical details of the calculations are presented in
the Appendix. Concluding remarks and a summary in sections \ref{Remarks} and
\ref{Summary} close the paper.

\section{The Model}
\label{model}

The ideal Boltzmann $\pi^+$ $\pi^-$ $\pi^0$ gas serves as the
standard example throughout this paper, while the main subject of
investigation is the multiplicity distribution $P(N_{\Omega})$ of particles
with momenta inside a certain segment $\Omega$ of momentum space. Calculations
are done for the three standard ensembles GCE, CE, and MCE. For the sake
of argument we will assume that we only want to measure  $P(N_{\Omega}^-)$,
i.e. the probability distribution of negatively charged pions in a limited
segment $\Omega$ of momentum space. Hence $\pi^-$ with momenta inside $\Omega$
are observed, while $\pi^-$ inside the complementary segment $\bar \Omega$ are
not observed. $\pi^+$ and $\pi^0$ are never detected. In GCE and CE the
presence of $\pi^0$ as a degree of freedom is of no relevance, while in MCE it
constitutes a heat bath for the remaining system. For consistency we use the
same system throughout this discussion. 

\begin{figure}[ht!]
\epsfig{file=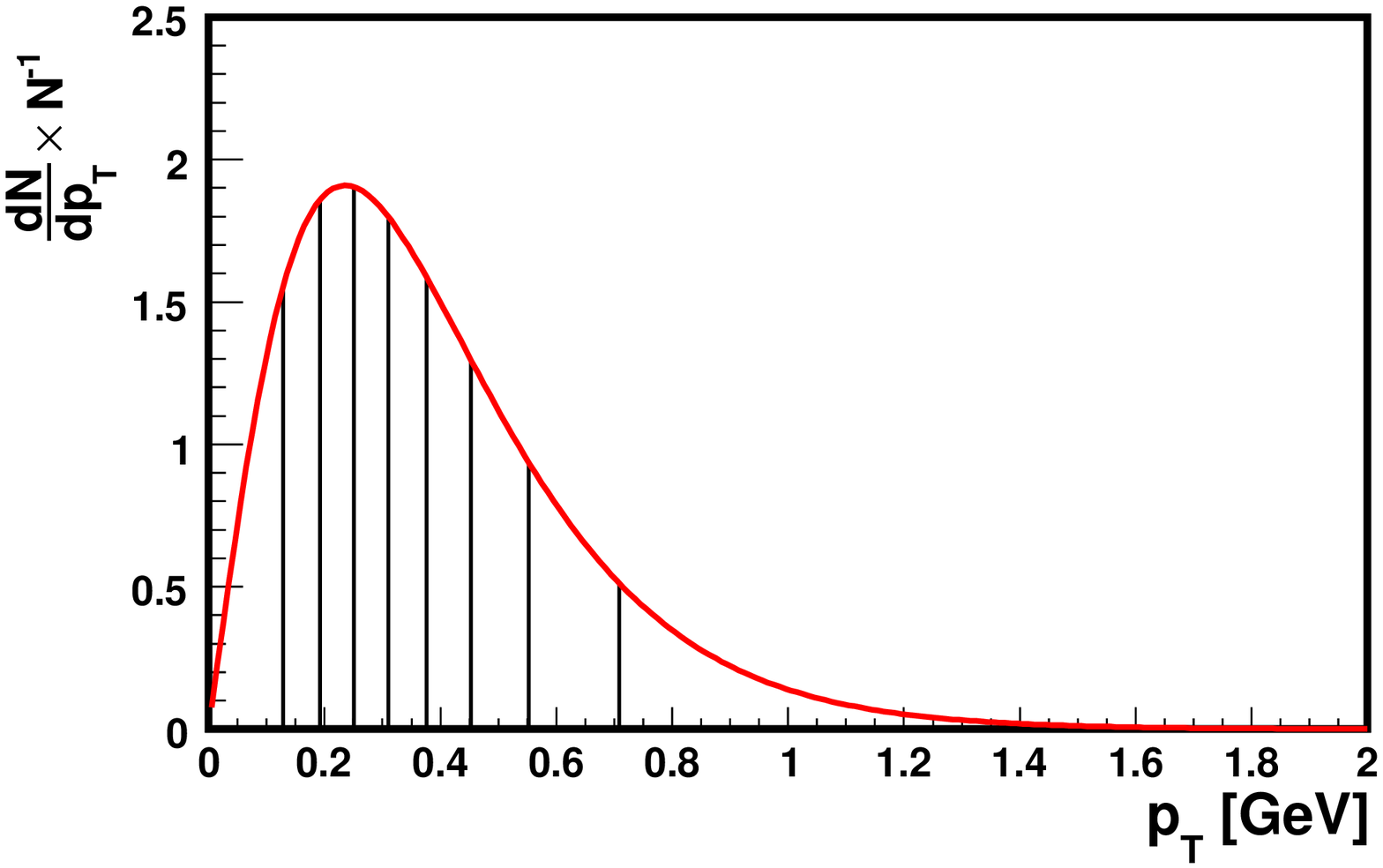,width=8cm}
\epsfig{file=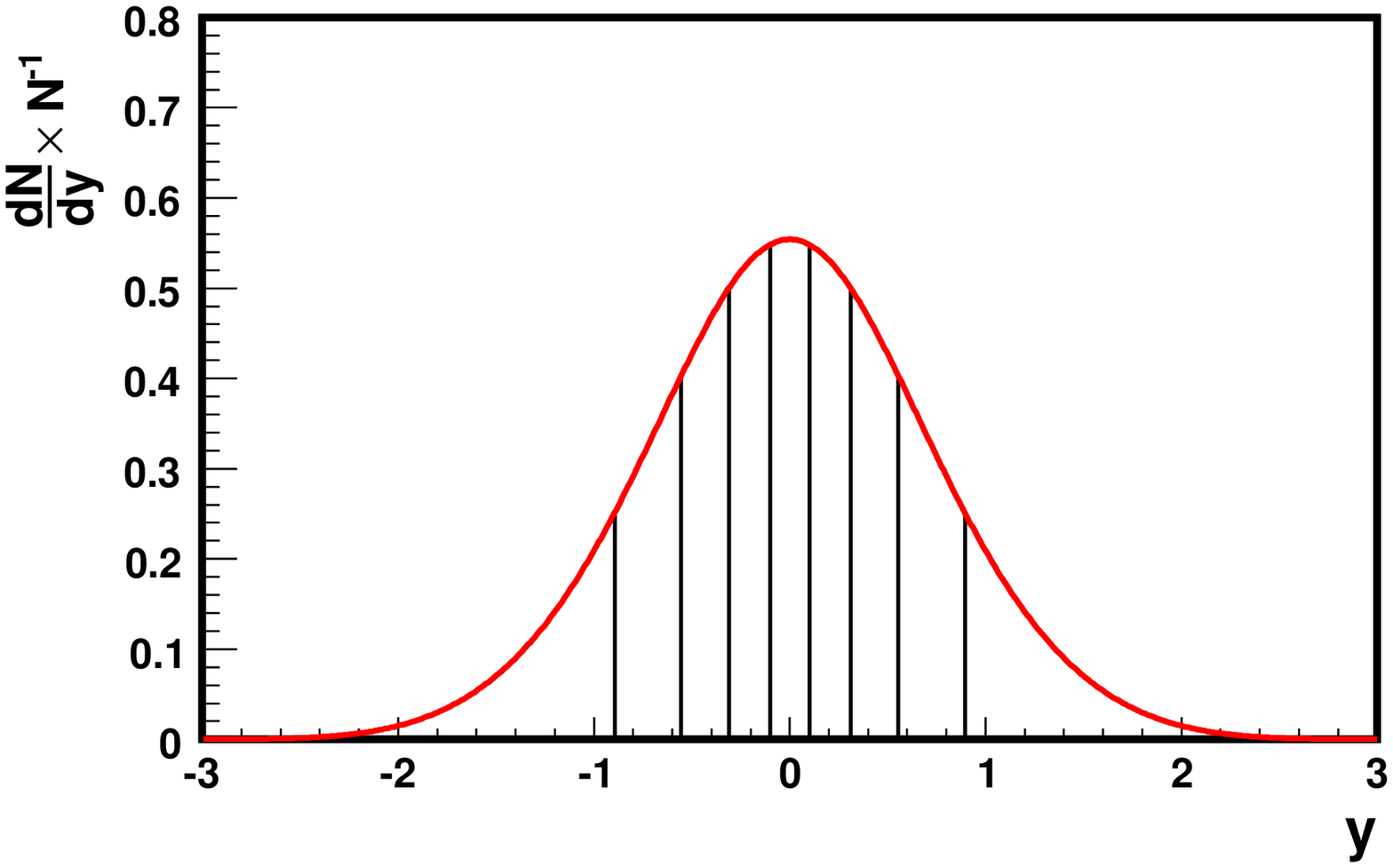,width=8cm}
\caption{Differential particle spectra for a classical pion gas at
  $T=160MeV$. Left: Transverse momentum spectrum Eq.(\ref{dNdpt}). Right:
  Rapidity spectrum Eq.(\ref{dNdy}). Both curves are normalized to the total
  yield Eq.(\ref{aveN}). The bins are constructed in a way that each bin
  contains   $1/9$ of the total yield.}  
\label{spectra}
\end{figure}

In order to keep the model simple, we assume a static homogenous
fireball. Our considerations therefore exclude collective motion, i.e. flow,
and resulting momentum spectra are purely thermal. We also omit resonance
decay contributions in this work. The spectra presented in Fig. \ref{spectra}
are normalized to the total $\pi^-$ yield in GCE and CE. Thus they are the
same in both ensembles. In MCE one expects  in the large volume limit only
small deviations from Boltzmann spectra. None of the forthcoming arguments
are affected by this. 
In the following we will use the transverse momentum and rapidity spectra
presented in Fig. \ref{spectra} to construct bins $\Omega_i = \Delta {p_T}_i =
\left[p_{T_i},p_{T_{i+1}} \right]$ (left), or $\Omega_i = \Delta y_i =
\left[y_i,y_{i+1} \right]$ (right), as indicates by the drop-lines.  

In section \ref{GCECE} we calculate the multiplicity distributions 
$P(N_{\Omega})$ for arbitrary segments $\Omega$ for the ideal Boltzmann GCE and
CE. To characterize the distribution one can calculate its (raw) moments
$\langle N_{\Omega}^n \rangle$ from:   
\begin{equation}\label{Moments}
\langle N_{\Omega}^n \rangle ~=~ \sum \limits_{N_{\Omega}=0}^{\infty}
~N_{\Omega}^n~ P \left(N_{\Omega} \right)~.
\end{equation}
A convenient measure for the width of a distribution is the scaled variance: 
\begin{equation}\label{ScaledVar} 
\omega_{\Omega} ~\equiv~ \frac{\langle N_{\Omega}^2 \rangle - \langle N_{\Omega}
  \rangle^2}{\langle N_{\Omega} \rangle} ~. 
\end{equation}  
In order to remove simple scaling effects, the bin sizes or segments are
chosen such that each bin or segment contains the same fraction $q= \langle
N_{\Omega} \rangle ~/~ \langle N_{4\pi} \rangle$ of the total yield (compare
Eq.(\ref{ScaledVar})). Here $\langle N_{\Omega} \rangle$ denotes the average
particle number in the momentum space segment $\Omega$, and  $\langle N_{4\pi}
\rangle$ denotes the average total ($4\pi$ integrated) multiplicity.  
The effect of finite acceptance can approximately be taken into account by
\cite{CEfluc1}:   
\begin{equation}\label{accscaling}
\omega_{q} ~=~ 1 + q \left(\omega_{4\pi}-1 \right)~,
\end{equation} 
where $\omega_{4\pi}$ assumes the ideal situation when all particles are 
detected, while $\omega_{q}$ assumes that particles are detected with 
probability $q$ regardless of their momentum. Hence Eq.(\ref{accscaling})
holds when particles are assumed to be uncorrelated in momentum space. In the 
limit $q\rightarrow 0$ one observes a random distribution with $\omega_q
\rightarrow 1$, i.e. a Poissonian, while when $q\rightarrow 1$ one sees the 
real distribution with width $\omega_q \rightarrow \omega_{4\pi}$. 
In this work we take explicitely correlations due to globally conserved 
charge (CE), and energy-momentum (MCE) into account and compare the results to
Eq.(\ref{accscaling}).

\section{Grand Canonical and Canonical Ensemble}
\label{GCECE}
\subsection{Grand Canonical Ensemble}
In the GCE, both, heat and charge bath are assumed to be infinite. And thus 
neither charge, energy nor momentum are conserved exactly. Temperature $T$ and
charge chemical potential $\mu$ regulate average energy and charge
density in a system of volume $V$. Usually it is said that charge, energy and
momentum are conserved in the average sense and fluctuations about an 
equilibrium value are allowed. Apart form Bose and Fermi effects \cite{Qstats}
particles are therefore uncorrelated in momentum space. However this example
serves as an illustration for the following CE and MCE calculations. We 
start by decomposing the Boltzmann single particle partition function
$z^-\left(\phi_{N_{\Omega}}\right)$ of $\pi^-$ into two parts,  
\begin{eqnarray}
z^-\left(\phi_{N_{\Omega}}\right)= z^-_{\Omega} \left(\phi_{N_{\Omega}}
\right) + z^-_{\bar \Omega} &=&  \frac{gV}{\left( 
    2\pi\right)^3} \int \limits_{\Omega} d^3 p ~ 
e^{-\frac{\varepsilon+\mu}{T}}~ e^{i \phi_{N_{\Omega}}} +
\frac{gV}{\left( 2\pi\right)^3} \int \limits_{\bar \Omega} d^3 p ~ 
e^{-\frac{\varepsilon+\mu}{T}} , 
\end{eqnarray}
where the single particle energy $\varepsilon = \sqrt{p^2+m^2}$, and $m$, and
$g$ are mass and degeneracy factor of $\pi^-$ respectively. Only for momentum
states inside the momentum space region $\Omega$ we introduce additionally a
Wick-rotated fugacity $\exp \left(i \phi_{N_{\Omega}} \right)$. For the
positive and neutral pion (which we do not want to detect in our example) we
write: 
\begin{eqnarray}
z^+ ~=~  \frac{gV}{\left( 2\pi\right)^3} \int  d^3 p ~
e^{-\frac{\varepsilon-\mu}{T}}~, 
\qquad \qquad \textrm{and} \qquad \qquad 
z^0 ~=~  \frac{gV}{\left( 2\pi\right)^3} \int  d^3 p ~
e^{-\frac{\varepsilon}{T}}~.
\end{eqnarray}
The value of the single particle partition function, for instance of the
neutral pion, is given by: 
\begin{equation}\label{aveN}
z^0=\langle N^0 \rangle = \frac{gV}{2\pi} m^2 T K_2 \left( \frac{m}{T}\right).
\end{equation}
For the sake of simplicity we assume equal masses for all pions. To obtain the
GCE multiplicity distribution for $N_{\Omega}$ in a momentum space segment
$\Omega$ we use the Fourier integral over the generalized GCE partition
function $ \mathcal{Z} \left( \phi_{N_{\Omega}}\right)=\exp \left[ z^-_{\Omega}
  \left( \phi_{N_{\Omega}} \right) + z^-_{\bar{ \Omega}}  + z^+ + z^0 \right] $,
normalized by the GCE partition function:
\begin{eqnarray}\label{GCEPDF}
P_{gce} \left(N_{\Omega} \right) ~\equiv~ Z^{-1}_{gce} \times \int
\limits_{-\pi}^{\pi} \frac{d\phi_{N_{\Omega}}}{2\pi} ~ e^{-iN_{\Omega}
  \phi_{N_{\Omega}} } ~ \mathcal{Z} \left( \phi_{N_{\Omega}}\right) ~=~
\frac{\left(z^-_{\Omega}\right)^{N_{\Omega}}}{N_{\Omega}!} 
\exp \left[- z^-_{\Omega} \right]~,
\end{eqnarray}
where the system partition function is given by $ Z_{gce} \equiv \mathcal{Z}
\left( \phi_{N_{\Omega}} = 0\right) $, and $z^-_{\Omega} =
z^-_{\Omega} \left(\phi_{N_{\Omega}}=0 \right)$. Independent of the 
shape or size of $\Omega$ we find a Poissonian for the multiplicity
distribution Eq.(\ref{GCEPDF}). Thus, using  Eq.(\ref{ScaledVar}), one finds
for the scaled variance $\omega^{gce}_{\Omega} = 1$, since $\langle
N_{\Omega} \rangle = z^-_{\Omega}$, and $\langle N^2_{\Omega}
\rangle = \langle N_{\Omega} \rangle^2 + \langle N_{\Omega} \rangle$. 

For Bose and Fermi statistics one does not expect a Poisson distribution and
(in particular when the chemical potential is large) deviations from a
Poissonian can be large. Thus one expects also deviations from
Eq.(\ref{accscaling}) when considering only finite acceptance.

\subsection{Canonical Ensemble}

In the CE the heat bath is still assumed to be infinite, while we remove the
charge bath and drop the chemical potential. Thus, we introduce a further 
Wick-rotated fugacity $\mu/T \rightarrow  i \phi_Q $ into the single particle
partition functions to account for global (however not in the momentum space
segment $\Omega$) conservation of electric charge $Q$. Particles in
$\Omega$ are therefore correlated, due to the condition of fixed net-charge,
with a finite charge bath composed of $\pi^+$ and  unobserved $\pi^-$. 
We again split the single particle partition function for $\pi^-$ into an
observed, $z^-_{\Omega}\left(\phi_{N_{\Omega}},\phi_Q\right)$, and an unobserved
part, $z^-_{\bar \Omega} \left(\phi_Q\right)$, 
\begin{eqnarray}
z^-\left(\phi_{N_{\Omega}},\phi_Q\right) = z^-_{\Omega}
\left(\phi_{N_{\Omega}},\phi_Q\right) + z^-_{\bar \Omega} 
\left(\phi_Q\right) =  \frac{gV}{\left(
    2\pi\right)^3} \int \limits_{\Omega} d^3 p ~ 
e^{-\frac{\varepsilon}{T}}  e^{-i \phi_Q} e^{i \phi_{N_{\Omega}}} +
\frac{gV}{\left( 2\pi\right)^3} 
\int \limits_{\bar \Omega} d^3 p ~ e^{-\frac{\varepsilon}{T}} e^{-i \phi_Q},
\end{eqnarray}
while we do not want to measure $\pi^+$ and $\pi^0$, and thus:
\begin{eqnarray}
z^+ \left(\phi_Q\right) ~=~  \frac{gV}{\left(
    2\pi\right)^3} \int  d^3 p ~ e^{-\frac{\varepsilon}{T}} e^{+ i \phi_Q}~,
\qquad \qquad \textrm{and} \qquad \qquad 
z^0 ~=~  \frac{gV}{\left(
    2\pi\right)^3} \int  d^3 p ~ e^{-\frac{\varepsilon}{T}} .
\end{eqnarray}
The normalization of the CE multiplicity distribution is given by the CE
system partition function $Z_{ce}$, i.e. the number of all micro states with
fixed charge Q,  $Z^{ce} = I_Q\left(2z \right) \exp(z^0)$, where $I_Q$ is the
modified Bessel function. The multiplicity distribution of $N_{\Omega}$ in
a momentum space segment $\Omega$, while charge $Q$ is globally conserved, can
be obtained from Fourier integration of the generalized GCE partition function
$\mathcal{Z} \left( \phi_{N_{\Omega}}, \phi_Q \right) = \exp \left[ z^-_{\Omega}
  \left( \phi_{N_{\Omega}},  \phi_Q \right)~+~ z^-_{\bar{ \Omega}}
  \left( \phi_Q \right) + z^+ \left(  \phi_Q \right) + z^0\right] $,
over both angles $\phi_Q$ and $\phi_{N_{\Omega}}$:
\begin{eqnarray}
P_{ce} \left(N_{\Omega}\right) &\equiv&  Z_{ce}^{-1} \times
\int \limits_{-\pi}^{\pi} \frac{d\phi_{N_{\Omega}}}{2\pi} \int
\limits_{-\pi}^{\pi} \frac{d\phi_{Q}}{2\pi} ~ e^{-iN_{\Omega}
  \phi_{N_{\Omega}} } ~ e^{-i Q \phi_{ Q} } ~\mathcal{Z} \left(
  \phi_{N_{\Omega}}, \phi_Q \right) \\ 
 &=& I_Q^{-1}\left(2z \right) \times 
\frac{\left(z^-_{\Omega}\right)^{N_{\Omega}}}{N_{\Omega}!} ~ \sum 
\limits_{a=0}^{\infty} ~ \frac{\left(z^-_{\bar \Omega}\right)^{a}}{a!} ~
\frac{z^{Q+N_{\Omega}+a}}{\left(Q+N_{\Omega}+a\right)!}~,
\end{eqnarray} 
where in CE $z^-_{\Omega} =  z^-_{\Omega} \left(\phi_{N_{\Omega}}=\phi_Q=0
\right)$,  $z^-_{\bar \Omega} =  z^-_{\bar \Omega} \left(\phi_Q=0 \right)$,
and $z= z^+\left(\phi_Q=0\right)=z^0$. For the respective first two moments
one finds from Eq.(\ref{Moments}):
\begin{eqnarray}
\langle N_{\Omega} \rangle = z^-_{\Omega } ~ \frac{I_{Q+1} \left(2z
  \right)}{I_Q   \left(2z \right)}~,  \qquad \textrm{and} \qquad 
\langle N^2_{\Omega} \rangle =   \left( z^-_{\Omega }\right)^2 ~\frac{I_{Q+2} 
  \left(2z \right)}{I_Q \left(2z \right)} +  z^-_{\Omega }~ \frac{I_{Q+1} 
  \left(2z\right)}{I_Q   \left(2z \right)}~. 
\end{eqnarray}
Thus, we obtain the well known canonical suppression of yields \cite{CEyield,
  CEfits, CEtransport, RateEqYield} and fluctuations \cite{CEfluc1,
  RateEqFluc}. The result, however, is completely independent of the
position of the segment $\Omega$. And therefore the scaled variance,
Eq.(\ref{ScaledVar}), takes the form:  
\begin{eqnarray}
\omega^{ce}_{\Omega} =  1 +  z^-_{\Omega} ~\left[ \frac{I_{Q+2}
  \left(2z\right)}{I_{Q+1}   \left(2z \right)}  -\frac{I_{Q+1}
  \left(2z\right)}{I_Q   \left(2z \right)}  \right]~, \qquad \textrm{and}
\qquad \omega^{ce}_{4\pi} =  1 +  z ~ \left[ \frac{I_{Q+2}
  \left(2z\right)}{I_{Q+1}   \left(2z \right)}  -\frac{I_{Q+1}
  \left(2z\right)}{I_Q   \left(2z \right)}  \right]~,
\end{eqnarray}
where $\omega_{\Omega}$ is the width of $P_{ce}(N_{\Omega})$, i.e. the
multiplicity distribution of $\pi^-$ with momenta inside $\Omega$, while
$\omega_{4\pi}$ is the width of the corresponding distribution when $\Omega$
is extended to the full momentum space. It can immediately be seen that this
formula is consistent with acceptance scaling, Eq.(\ref{accscaling}),
$\omega_{\Omega} ~=~ 1 + q \left(\omega_{4\pi} -1 \right)$,  if $q \equiv
z^-_{\Omega}/z$. Generally we find $\omega^{ce}_{4\pi} < \omega^{ce}_{\Omega}
< \omega^{gce}=1$. In the limit of $z^-_{\Omega}/z \rightarrow 0$ we approach
the Poisson limit of a `random` distribution with $\omega = 1$, i.e. the
observed part of the system is embedded into a much larger charge bath and the
GCE is a valid description.   

\section{Micro-Canonical Ensemble} 
\label{MCE}
For the MCE an analytical solution seems to be out of reach presently, so we
use instead the asymptotic solution, applicable to large systems, derived in
\cite{clt}. In order to avoid unnecessary repetition of calculations, we will
only give a general outline here, and refer the reader for a detailed
discussion to Ref.\cite{clt}. It should be mentioned that this method
be would be also applicable to systems of finite spatial extention, provided
the average particle number in a given momentum space bin exceeds roughly
$\langle N_{\Omega}\rangle \gtrsim 5$. In this work we confine ourselves to
large systems and try to asses the general trends. 

The basic idea is to define the MCE multiplicity distribution in terms of a
joint GCE distribution of multiplicity, charge, energy, momentum, etc. The
MCE multiplicity distribution is then given by the (normalized) conditional 
probability in the GCE to find a number $N_{\Omega}$ of particles in a segment
$\Omega$ of momentum space, while electric charge $Q$, energy $E$, and three
momentum $\vec P$ are fixed. Therefore we will keep temperature and chemical
potentials as  parameters to describe our system. Effective temperature and
effective chemical potential, i.e. Lagrange multipliers, can be determined by
demanding that the GCE partition function is maximized for a certain
equilibrium state $(Q,E,\vec P)$. This requirement is entirely consistent
\cite{clt} with the usual textbook definitions of $T$ and $\mu$ in MCE and CE
through differentiation of entropy and Helmholtz free energy with respect to
conserved quantities. In principle we would have to treat all conservation
laws on equal footing \cite{MCEmagic}, and thus introduce Lagrange multipliers
for momentum conservation as well. However here we are only interested in a
static source, thus $\vec P = \vec 0$, and the relevant parameters are equal
to zero.  

In the large volume limit energy, charge, and particle density in the MCE will
correspond to GCE values. This is required by the  thermodynamic equivalence
of ensembles for mean quantities. MCE and CE partition functions are generally 
obtained from their GCE counterpart by multiplication with delta-functions, 
which pick out a set of micro states consistent with a particular
conservation law. Here it will be of considerable advantage to use Fourier
representations of delta-functions, similar to the treatment in Section
\ref{GCECE}. This method could be considered to be a Fourier spectral analysis
of the generalized GCE partition function \cite{clt}. 
 
The normalized conditional probability distribution of multiplicity $N_{\Omega}$
can be defined by the ratio of the values of two partition functions:
\begin{equation}\label{MCEprob}
P_{mce}(N_{\Omega})~\equiv~ \frac{\textrm{number of all states with $N_{\Omega}$,
    $Q$, $E$, and $\vec P = \vec 0$}}{\textrm{number of all states with $Q$,
    $E$, and  $\vec P= \vec 0$} }~.
\end{equation}
The real MCE partition function and our modified version are connected as
$Z(V,N_{\Omega},Q,E,\vec P)  \equiv  \mathcal{Z}^{N_{\Omega},Q,E,\vec
  P}(V,T,\mu) e^{+E/T} e^{-Q\mu/T}$. In either case the normalization in
Eq.(\ref{MCEprob}) is given by the partition functions with fixed values of
$Q,E,\vec P$, but arbitrary particle number $N_{\Omega}$, hence $Z(V,Q,E,\vec
P) \equiv \sum_{N_{\Omega}=0}^{\infty} Z(V,N_{\Omega},Q,E,\vec P)$, or
$\mathcal{Z}^{Q,E,\vec P}(V,T,\mu) \equiv \sum_{N_{\Omega}=0}^{\infty}
\mathcal{Z}^{N_{\Omega},Q,E,\vec  P}(V,T,\mu)$. However when taking the ratio
(\ref{MCEprob}) auxiliary parameters chemical potential and temperature 
drop out:  
\begin{equation}\label{MCEprob2}
P_{mce}(N_{\Omega})~ \equiv~ \frac{Z(V,N_{\Omega},Q,E,\vec P)}{Z(V,Q,E,\vec
  P)} ~=~ \frac{\mathcal{Z}^{N_{\Omega},Q,E,\vec
    P}(V,T,\mu)}{\mathcal{Z}^{Q,E,\vec P}(V,T,\mu)}~.
\end{equation}
The main difference between the two versions of partition functions is that for
$Z(V,N_{\Omega},Q,E,\vec P)$ one is confronted with a heavily oscillating
(or even irregular) integrant, while for $\mathcal{Z}^{N_{\Omega},Q,E,\vec
  P}(V,T,\mu)$ the integrant becomes ($T$,$\mu$ correctly chosen) very
smooth. Thus, introduction of $T$ and $\mu$ allows to derive (and use) the
asymptotic solution of Ref.\cite{clt}.  

We have a total number of 6 conserved `charges`, and hence we need to solve
the 6-dimensional Fourier integral for the numerator in
Eq.(\ref{MCEprob2})\footnote{We drop in the following the argument $(V,T,\mu)$
  to simplify the notation.}:   
\begin{eqnarray}\label{MCEInt}
\mathcal{Z}^{N_{\Omega},Q,E, \vec P}&=&  
\int \limits_{-\pi}^{\pi} \frac{d \phi_{N_{\Omega}}}{2\pi}  
\int \limits_{-\pi}^{\pi} \frac{d \phi_Q}{2\pi}  
\int \limits_{-\infty}^{\infty} \frac{d \phi_E}{2\pi}  
\int \limits_{-\infty}^{\infty} \frac{d \phi_{P_x}}{2\pi}  
\int \limits_{-\infty}^{\infty} \frac{d \phi_{P_y}}{2\pi}  
\int \limits_{-\infty}^{\infty} \frac{d \phi_{P_z}}{2\pi} \nonumber \\ 
&\times& e^{-iN_{\Omega} \phi_{N_{\Omega}}}~ e^{-iQ\phi_Q} ~e^{- iE \phi_E} 
~e^{-iP_x \phi_{P_x}}~ e^{-iP_y \phi_{P_y}}~ e^{-iP_z \phi_{P_z}}   \nonumber \\ 
&\times& \exp \left[  V \sum_k \psi_k \left( \phi_{N_{\Omega}}\phi_Q, \phi_E,
\phi_{P_x},\phi_{P_y},\phi_{P_z} \right) \right].
\end{eqnarray}
The summation in (\ref{MCEInt}) should be taken over the single particle
partitions $V \psi_k=z_k$ of all considered particle species $k$. The
Wick-rotated fugacities $\phi_{Q}$, etc. are related to the
individual conservation laws. The distinction between the Kronecker
delta-function (limits of integration $\left[-\pi,\pi \right]$) for discrete
quantities and the Dirac delta-function (limits of integration
$\left[-\infty,\infty \right]$) for continuous quantities is important here,
however for deriving an asymptotic solution it will not be. To simplify 
(\ref{MCEInt}) we change to shorthand notation for $\phi_j =
(\phi_{N_{\Omega}}\phi_Q, \phi_E, \vec \phi_P)$ and the conserved `charge`
vector $ Q^j = (N_{\Omega},Q,E,\vec P)$. We again split the single particle
partition  functions in two parts. The first part counts the number of
momentum states observable to our detector, while the second part counts
momentum states invisible to our detector:
\begin{eqnarray}
\psi_{k} \left( \phi_j\right)  &=&  \frac{g_k}{\left( 2\pi\right)^3} \int
\limits_{ \Omega} d^3 p ~ e^{-\frac{\varepsilon_k - q_k \mu}{T}} ~e^{i q_{k,
    \Omega}^j \phi_j}~ +~ \frac{g_k}{\left( 2\pi\right)^3} 
\int \limits_{ \bar \Omega} d^3 p~ e^{-\frac{\varepsilon_k-q_k \mu}{T}} ~ e^{i
  q_{k,\bar \Omega}^j \phi_j}~.
\end{eqnarray}
For the `charge` vector of all measured particle species $k$ we write
$q^j_{k,\Omega} = (1,q_k,\varepsilon_k,\vec p_k)$ for momenta inside $\Omega$,
and $q^j_{k, \bar \Omega} = (0,q_k,\varepsilon_k,\vec p_k)$ for momenta
outside of $\Omega$. For all unobserved particle species we write
$q^j_{k,\Omega}=q^j_{k, \bar \Omega} =(0,q_k,\varepsilon_k,\vec p_k) $. Here
$q_k$ is the electrical charge of particle species $k$, and $\varepsilon_k$ and
$\vec p_k$ are its energy and momentum vector. In Ref.\cite{clt}, where only
multiplicity distributions in the full momentum space were considered, the
general `charge` vector took the form $q^j_{k,4\pi} =
(n_k,q_k,\varepsilon_k,\vec p_k)$, where $n_k$ is the multiplicity of this
particle. For stable particles $n_k=1$ in case they are observed, and $n_k=0$
if they are not measured, while for unstable particles $n_k$ could also denote
the number of measurable decay products. 

For large system volume the main contribution to the integral (\ref{MCEInt})
comes from a small region around the origin \cite{VolDep}. Thus we proceed
by Taylor expansion of the integrant of (\ref{MCEInt}) around $\phi_j=\vec
0$. In this context  $ \Psi \left( \phi_j\right) = \sum_k \psi_k \left(
  \phi_j\right) $ would be called the cumulant generating function
(CGF). Cumulants (expansion terms) are defined by differentiation of the
CGF at the origin: 
\begin{equation}\label{kappa_n}
\kappa_n^{j_1,j_2,\dots,j_n } ~\equiv~ \left(-i\right)^n\frac{\partial^n
\Psi \left( \phi_j \right) }{\partial \phi_{j_1}
  \partial \phi_{j_2} \dots \partial \phi_{j_n} }
\Bigg|_{\phi_j = \vec 0}~.
\end{equation}
Generally are cumulants tensors of rank $n$ and dimension defined by the
number of conserved quantities. Here $\kappa_1$ is a 6 component vector, while 
$\kappa_2$ is a $6 \times 6$ matrix, etc.

The parts of the integrant related to discrete quantities, i.e. $N_{\Omega}$
and $Q$, are now not $2\pi$ periodic anymore (while in Eq.(\ref{MCEInt}) they
are), but superpositions of oscillating and decaying parts. Thus we extent the
limits of integration to $\pm \infty$, what introduces a negligible
error. Eq.(\ref{MCEInt}) therefore simplifies to:   
\begin{eqnarray}\label{MCEIntApprox}
\mathcal{Z}^{Q^j} &\simeq&   \left[ \prod_{j=1}^{6} \int
  \limits_{-\infty}^{\infty} \frac{d\phi_j}{\left( 2\pi \right)} \right] ~\exp
\Big[  -iQ^j\phi_j
~+~V \sum_{n=0}^{\infty} \frac{i^n}{n!} \; \kappa_n^{j_1,j_2,\dots,j_n } \;
\phi_{j_1} \phi_{j_2} \dots \phi_{j_n} \Big] ~.
\end{eqnarray}
Summation over repeated indices is implied. Existence and finiteness of the
first three cumulants provided, any such integral can be shown to
converge to a multivariate normal distribution in the large volume limit:
\begin{equation}\label{MCE_MultNormal}
\mathcal{Z}^{Q^j} ~\simeq~ Z_{gce} \frac{\exp \left(-\frac{\xi^j \; \xi_{ j}}{2}
  \right)}{\left(2\pi V \right)^{6/2} \det| \sigma| }~,
\end{equation}
where $Z_{gce}\equiv \exp \left[V \kappa_0 \right]$ is the GCE partition
function, $\kappa_0$ is the cumulant of $0^{th}$ order, $\xi^j=\left( Q^k - V
  \kappa_1^k \right) \left( \sigma^{-1}\right)_{k}^{\;\;j}  V^{-1/2}~$ is a
measure for the distance of a particular macro state $Q^k$ to the peak $V
\kappa_1^k$ of the joint distribution, and $\sigma$ is the square root  of the
second rank tensor $\kappa_2$, see \cite{clt} for details. 

The normalization in Eq.(\ref{MCEprob2}) can essentially be found in two
ways. The first way would be to integrate the distribution
(\ref{MCE_MultNormal}) over all possible values of multiplicity $N_{\Omega}$,
while all other variables are set to their peak values, e.g. $Q=V\kappa_1^Q$,
$E=V\kappa_1^E$, $\vec P = \vec 0$. The second and more practical way is to use
an approximation similar to Eq.(\ref{MCE_MultNormal}) to describe the macro
state $Q^j = (Q,E,\vec P)$. The normalization in Eq.(\ref{MCEprob2}),
$\mathcal{Z}^{E,Q,\vec P}$, is then given by the 5-dimensional integral,
similar to Eq.(\ref{MCEInt}), without the integration over
$\phi_{N_{\Omega}}$. The 1-dimensional slice along $N_{\Omega}$, i.e. the 
conditional distribution of particle number $N_{\Omega}$, while charge, energy
and momentum are fixed to $Q,E,\vec P = \vec 0$, can then be shown \cite{clt}
to converge to a Gaussian in the large volume limit: 
\begin{eqnarray}\label{PMCE}
P_{mce}(N_{\Omega} ) ~ \simeq~  \frac{1}{\left(2\pi ~\omega^{mce}_{\Omega}
   ~ \langle N_{\Omega} \rangle   \right)^{1/2}} ~  \exp  \left(- \frac{ \left(
      N_{\Omega} - \langle N_{\Omega} \rangle \right)^2}{2
    ~  \omega^{mce}_{\Omega} ~ \langle  N_{\Omega} \rangle  } \right) ~. 
\end{eqnarray}
The scaled variance $\omega^{mce}_{\Omega}$ is given by the ratio of the two
determinants of the two relevant second rank cumulants, $\kappa_2$ and $\tilde
\kappa_2$, of the two partition functions $\mathcal{Z}^{N_{\Omega},E,Q,\vec
  P}$ and  $\mathcal{Z}^{E,Q,\vec P}$, hence\footnote{Please note, that in
  order to simplify formulas, the notation is slightly different from
  \cite{clt}. }:  
\begin{equation}\label{SimpleOmega}
\omega^{mce}_{\Omega} =  \frac{ \det | \kappa_2| }{
  \kappa_1^{N_{\Omega}} \;\det| \tilde \kappa_2|  }~. 
\end{equation}
The asymptotic ($V\rightarrow \infty$) scaled variance can therefore be
written in the form of Eq.(28) in \cite{clt}. Considering only the asymptotic
solution we need to investigate only the first two cumulants ($n=1,2$) in
detail. We will first discuss the structure of $\kappa_1$ and $\kappa_2$, and
then deduce a few properties of Eq.(\ref{SimpleOmega}).

The first order cumulant $\kappa_1$ of $\mathcal{Z}^{N_{\Omega},Q,E,\vec
  P}$ gives GCE expectation values for particle density
$\kappa_1^{N_{\Omega}}$, charge density $\kappa_1^{Q}$, energy density
$\kappa_1^{E}$, and expectation values of momentum 
$\kappa_1^{p_x}$, etc. Since we are only interested in a static source we find
due to the antisymmetric momentum integral (see Appendix \ref{Calc})
$\kappa_1^{p_x} = \kappa_1^{p_y} = \kappa_1^{p_z}=0$.  The general form of the
first cumulant $\kappa_1$ is then: 
\begin{align}
\kappa_1 = 
\begin{pmatrix}
\kappa_1^{N_{\Omega}}, & \kappa_1^{Q}, & \kappa_1^{E}, & 0, & 0, &  0 
\end{pmatrix}~. \label{vector}
\end{align} 

The second cumulant $\kappa_2$ of $\mathcal{Z}^{N_{\Omega},Q,E,\vec
  P}$ contains information about correlations due to different conserved
quantities. A detailed discussion of
correlation terms only involving Abelian charges and/or energy,
e.g. $\kappa_2^{Q,Q}$, $\kappa_2^{Q,E}$, and $\kappa_2^{E,E}$, can be found in
\cite{clt}. Again, due to the antisymmetric nature of the momentum integral,
all cumulant entries involving an odd order in one of the momenta,
e.g. $\kappa_2^{E,p_x}$, $\kappa_2^{p_x,p_y}$, or $\kappa_2^{Q,p_x}$ are equal to
zero. The general second order cumulant $\kappa_2$ thus reads:  
\begin{align}
\kappa_2 = 
\begin{pmatrix}
\kappa_2^{N_{\Omega},N_{\Omega}} & \kappa_2^{N_{\Omega},Q} &
\kappa_2^{N_{\Omega},E} & \kappa_2^{N_{\Omega},p_x} &
\kappa_2^{N_{\Omega},p_y} &  \kappa_2^{N_{\Omega},p_z} \\ 
\kappa_2^{Q,N_{\Omega}} & \kappa_2^{Q,Q} & \kappa_2^{Q,E} & 0 & 0 &  0 \\ 
\kappa_2^{E,N_{\Omega}} & \kappa_2^{E,Q} & \kappa_2^{E,E} & 0 & 0 & 0 \\ 
\kappa_2^{p_x,N_{\Omega}} & 0 & 0 & \kappa_2^{p_x,p_x} & 0 & 0 \\ 
\kappa_2^{p_y,N_{\Omega}} & 0 & 0 & 0 & \kappa_2^{p_y,p_y} & 0 \\ 
\kappa_2^{p_z,N_{\Omega}} & 0 & 0 & 0 & 0 &  \kappa_2^{p_z,p_z} 
\end{pmatrix}~. \label{matrix}
\end{align}
Please note that by construction, Eq.(\ref{kappa_n}), the matrix
(\ref{matrix}) is symmetric, hence $\kappa_2^{N_{\Omega},Q} =
\kappa_2^{Q,N_{\Omega}}$, etc. 

The second matrix $\tilde \kappa_2$, now related to the partition function
$\mathcal{Z}^{Q,E,\vec  P}$, is obtained from $\kappa_2$, Eq.(\ref{matrix}), by
crossing out the first row and first column. In the following we are going to
make use of the fact that one can express the  determinant of a matrix $A$ by:
\begin{equation}\label{calcdet}
\det |A| ~=~ \sum \limits_{j=1}^n \left( -1\right)^{j+k}
A_{j,k}  M_{j,k} ~,
\end{equation}
where $A_{j,k}$ is the matrix element $j,k$ of a general non-singular $n\times
n$ matrix $A$, and $ M_{j,k}$ is its complementary minor. A simple consequence
of Eq.(\ref{calcdet}) is:  
\begin{equation}\label{normdet}
\det |\tilde \kappa_2| = \kappa_2^{p_x,p_x} ~\kappa_2^{p_y,p_y} ~\kappa_2^{p_z,p_z}
\left[ \kappa_2^{E,E}~\kappa_2^{Q,Q}- \left(\kappa_2^{E,Q}\right)^2 \right] ~=~
\left(  \kappa_2^{p_x,p_x}  \right)^3 \det |\hat \kappa_2|,
\end{equation}
where $\kappa_2^{p_x,p_x} =\kappa_2^{p_y,p_y} =\kappa_2^{p_z,p_z} $, due to
spherical symmetry in momentum space, and $\hat \kappa_2$ is just a $2\times2$
matrix involving only terms containing $E$ and $Q$. In case correlations
between particle number and conserved momenta are vanishing,
i.e. $\kappa_2^{N_{4\pi},p_x} = 0$, or $\kappa_2^{N_{\Omega},p_x} = 0$, then,
similarly to Eq.(\ref{normdet}), the determinant of $\kappa_2$ factorizes into
a product of correlation terms $(\kappa_2^{p_x,p_x})^3$ and the determinant of
a $3\times3$  sub-matrix involving only terms containing $E$, $Q$, and
$N$. Hence in taking the ratio Eq.(\ref{SimpleOmega}) one notes, that in this
case momentum conservation will not affect multiplicity fluctuations  in the
large volume limit \cite{clt}. In this work, however we do not
necessarily find $ \kappa_2^{N_{\Omega},p_x} = 0 $, as we only integrate over
a limited segment $\Omega$ of momentum space, and taking momentum conservation
into account may affect the result. 

Finally it should be stressed that this procedure can be easily generalized to
account for Bose or Fermi statistics. Also phenomenological  phase space
suppression (enhancement) factors $\gamma_q$ \cite{gammaQfirst} or $\gamma_s$
\cite{gammaSfirst} could be straightforwardly included. However, without
proper implementation of the effect of additional correlations due to
resonance decay and collective motion, i.e. flow, it seems of little value to
do too strict calculations for experimentally measurable distributions. We
thus return to the pion gas example from section \ref{GCECE} and restrict the
discussion to simple momentum space cuts in rapidity, transverse momentum, and
azimuthal angle, see also the Appendix for details. 

\section{Results}  
\label{Results} 
\subsection{Multiplicity fluctuations in the full momentum space}
Let us firstly recall basic properties of multiplicity fluctuations of
negative particles in the full momentum space ($4\pi$ fluctuations) in the
three standard ensembles, of the Boltzmann pion gas considered
here. 

Multiplicity fluctuations in the CE are suppressed due to exact charge
conservation. For a neutral ($Q=0$) system one finds in the large volume limit 
$\omega_{4\pi}^{ce} = 0.5$ \cite{CEfluc1}. Further suppression of 
fluctuations arise from additionally enforcing exact energy conservation in
the MCE. Here one finds $\omega_{4\pi}^{mce} \approx 0.25$ for a Boltzmann
pion gas at $T\approx 160MeV$. In the GCE, since no conservation laws are
enforced, we always find a Poisson distribution with width
$\omega_{4\pi}^{gce} =1$.  

Since charge conservation in CE links the distributions of negatively charged
particles to the one of their positive counterparts, i.e. $P(N_-) = P(N_+-Q)$,
the relative width  of $P(N_-)$ increases (decreases) as we move the electric
charge density to positive (negative) values \cite{CEfluc2}. This can be
easily be seen from Eq.(\ref{matrix}) by crossing out all rows and columns
containing energy and momentum  and calculating the asymptotic scaled variance
of negatively charged particles, $\omega^{ce}_{4\pi}$, from
Eq.(\ref{SimpleOmega}),    
\begin{equation}
\omega_{4\pi}^{ce}~=~ \frac{\kappa_2^{N_{4\pi},N_{4\pi}}\kappa_2^{Q,Q}-\left( 
    \kappa_2^{N_{4\pi},Q} \right)^2}{\kappa_1^{N_{4\pi}}~ \kappa_2^{Q,Q}} ~=~
\frac{\exp \left( \frac{\mu}{T} \right)}{2\cosh \left(
    \frac{\mu}{T}\right)}~. 
\end{equation}
The same effect is present in the MCE, however the calculation is slightly
longer. 

Results for $4\pi$ multiplicity fluctuations of negatively charged particles
in a Boltzmann pion gas at $T=160MeV$ and different charge densities are 
summarized in Table \ref{table}. Additionally estimates, based on our
previously employed `uncorrelated particle` approach, Eq(\ref{accscaling}),
for multiplicity  fluctuations with limited acceptance are given.    
\begin{table}[h!]
\begin{center}
\begin{tabular}{c||c|c|c||c|c|c}
& $\quad \omega^{gce}_{4\pi} \quad$ & $\quad \omega^{ce}_{4\pi} \quad$ & $ 
\quad \omega^{mce}_{4\pi} \quad$ & $ \quad \omega^{gce}_{q=1/9} \quad$ &
$\quad \omega^{ce}_{q=1/9} \quad$ & $\quad \omega^{mce}_{q=1/9} \quad$ \\ 
\hline \hline
$\mu=0$            & $1$ & $0.5$   & $0.235$ & $1$ & $0.944$ & $0.915$\\ 
$\mu=-\frac{m}{2}$ & $1$ & $0.294$ & $0.147$ & $1$ & $0.922$ & $0.905$\\
$\mu=+\frac{m}{2}$ & $1$ & $0.706$ & $0.353$ & $1$ & $0.967$ & $0.928$\\ 
 \end{tabular}
 \caption{Multiplicity fluctuation of $\pi^-$ in a classical pion gas in the 
   large volume limit in the three standard ensembles at $T=160MeV$ for
   different charge densities. The index `$4\pi$` denotes fluctuations in the 
   full momentum space, while the index `$q=1/9$` assumes acceptance scaling,
   Eq.(\ref{accscaling}). The ratio $n_-/n_{tot}$ equals to $0.33$ for 
   $\mu=0$, $0.48$ for $\mu=-m/2$, and $0.20$ for $\mu=+m/2$.}   
 \label{table} 
\end{center} 
\end{table}
Despite the fact that $\omega_{4\pi}$ is very different in GCE, CE, or MCE and 
also rather sensitive to the charge density, the estimates for limited
acceptance ($q=1/9$) based on Eq.(\ref{accscaling}) vary only by a few \%. In
order to decisively distinguish predictions for different ensembles a large
value of $q$ would be needed. 

\subsection{Multiplicity fluctuations in limited segments of momentum space}

In Section \ref{GCECE} we have seen that in the Boltzmann CE multiplicity
fluctuations observed in a limited segment of phase space are insensitive to
the position of this segment. The dependence on the size of the segment can
thus be taken into account by use of acceptance scaling
Eq.(\ref{accscaling}). To balance charge a particle can be produced or
annihilated anywhere in momentum space. And due to a infinitely large heat and
momentum bath in the CE no momentum state is essentially preferred.  

In the MCE this dependence is qualitatively different. When using the
MCE formulation particles are correlated due to the constraints of exactly
conserved energy and momentum, even in the large volume limit. Fluctuations in
a macroscopic subsystem are strongly  affected by correlations with the
remainder of the system. 

\begin{figure}[ht!]
\epsfig{file=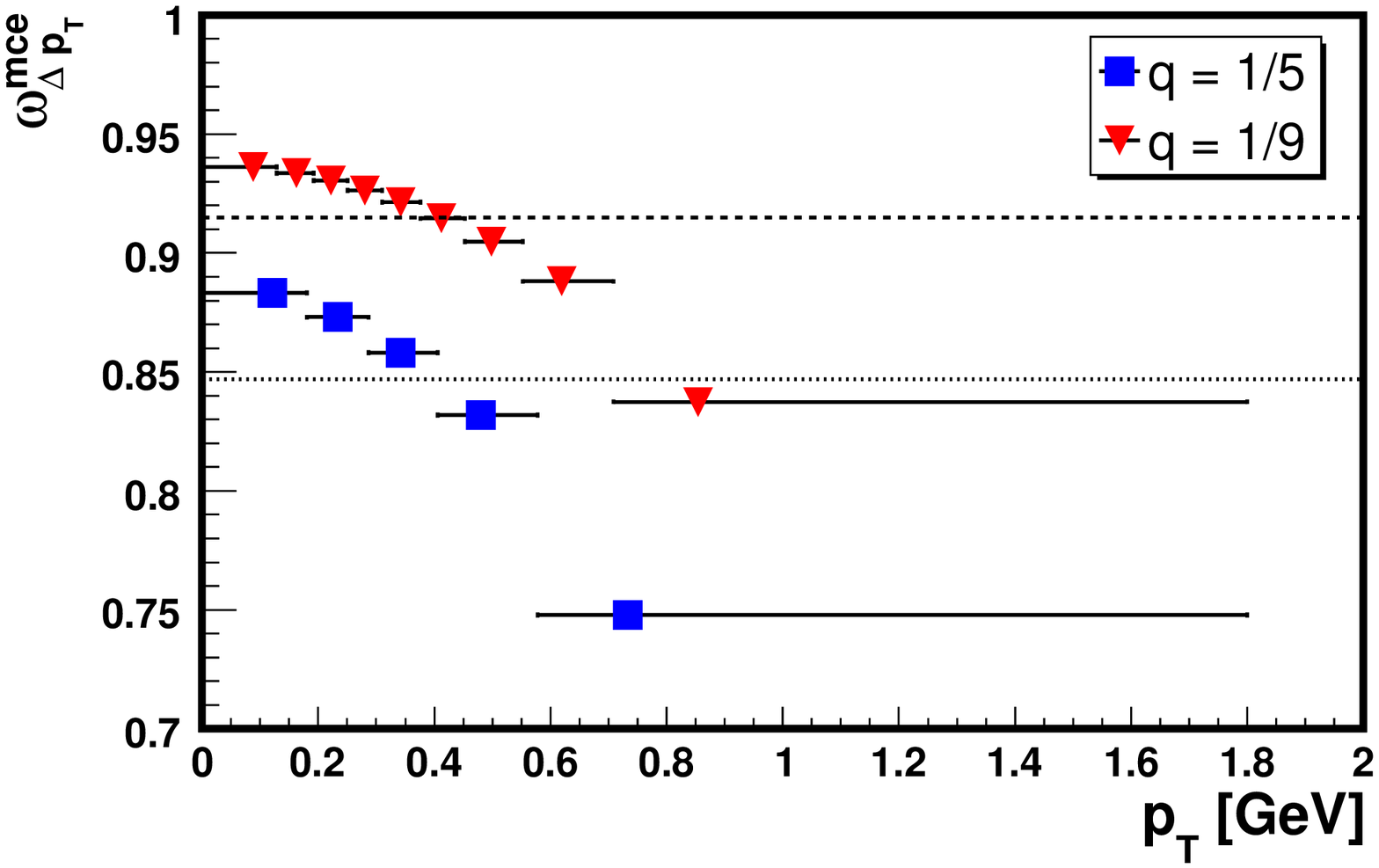,width=8cm}
\epsfig{file=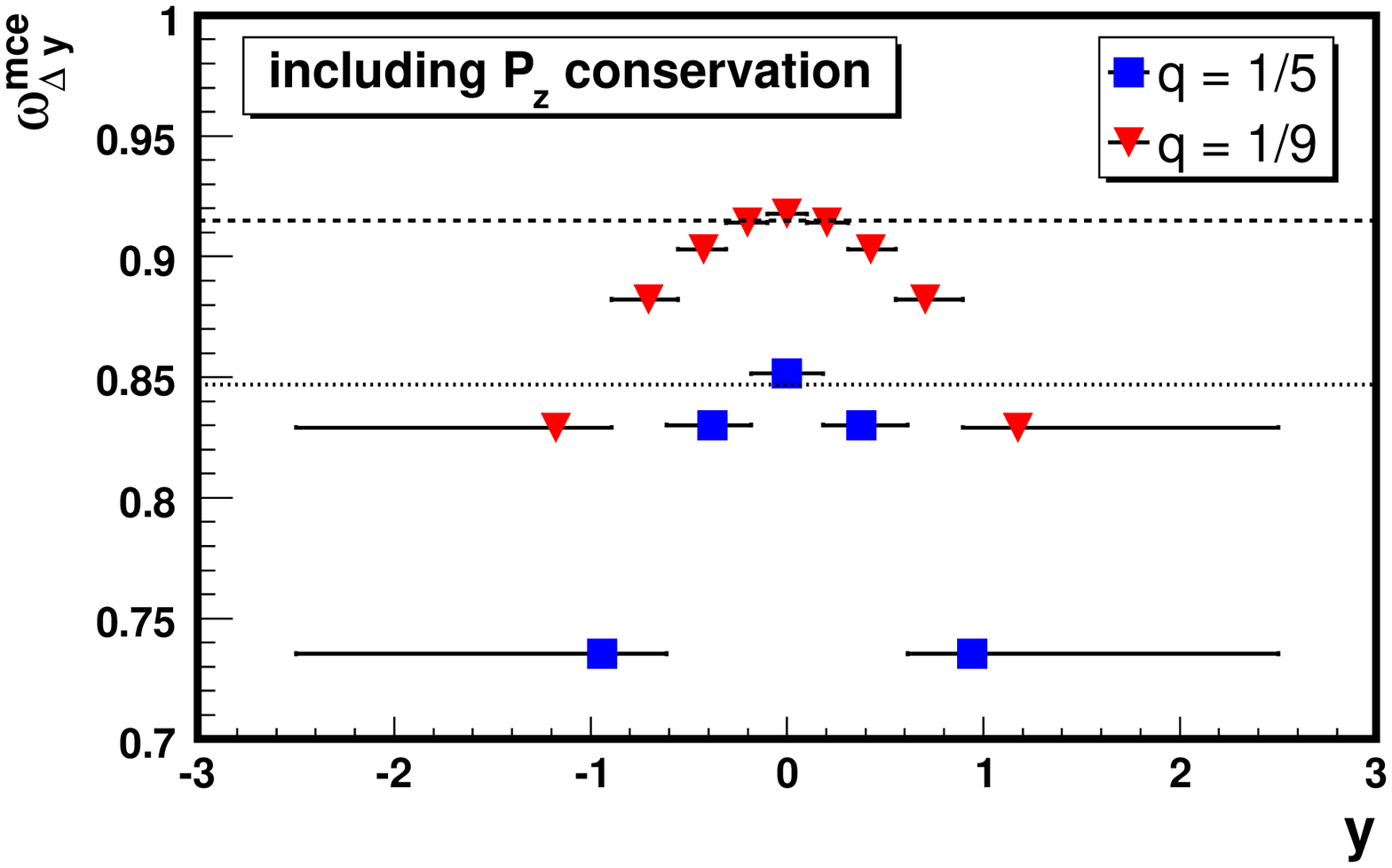,width=8cm}
\caption{Transverse momentum (left) and rapidity dependence (right) of the
  scaled variance of $\pi^-$ at $T=160MeV$, for a classical pion gas at zero
  charge density. Momentum bins are constructed in a way that each bin
  contains the same fraction $q$ of the average $\pi^-$ yield. The (error)bars
  indicate the width of the $p_T$ or $y$ bins,  while the marker indicate the
  center of gravity of the corresponding bin. The lines indicate acceptance
  scaling  Eq.(\ref{accscaling}). Calculations are done for different values of
  acceptance. $q=1/5$ (square marker, dotted line), $q=1/9$ (triangle down,
  dashed).}  
\label{dodp}
\end{figure}

In Fig. \ref{dodp} we show the scaled variance of multiplicity fluctuations
for negatively charges particles in finite bins in transverse momentum (left),
and rapidity (right). The bins are constructed such that each bin contains on
average the same fraction $q$ of the total average yield. The width of each
bin is indicated by the bars. Calculations are done for two values of
acceptance ($q=1/5$, and $q=1/9$). The dashed and dotted lines correspond to
acceptance scaling Eq.(\ref{accscaling}), while the markers are calculated from
Eq.(\ref{SimpleOmega}). One finds that multiplicity fluctuations in bins with
high transverse momentum and high values of rapidity are, due to energy and
momentum conservation, essentially suppressed with respect to bins where
individual particles carry less energy and momentum.  

A intuitive explanation would probably look like this: Let us consider an
event with an unusually large (small) number of  particles at the most forward
rapidity bin. In this bin we would find therefore a macroscopic state with
unusually large (small) observed longitudinal momentum $P^{obs}_z$ and energy  
$E^{obs}$. The remainder of the system therefore has to have rather large
(small) momentum $-P_z^{obs}$ and rather small (large) energy $E-E^{obs}$.
Since both probability distributions, for the observed and the
unobserved subsystems, do not factorize into independent probability
distributions, but are correlated, this macro state would be rather
unlikely. Fluctuations about the mean $\langle N_y \rangle$ at forward
(backward) rapidities should therefore be suppressed. On the other hand can
modest multiplicity fluctuations in a high $p_T$ bin induce stronger
fluctuations in the lower $p_T$ bins, and fluctuations about $\langle N_{p_T}
\rangle$ in a low $p_T$ bin are enhanced. Even when detecting only a fraction
of about 10\% of the total system these correlations can have a sizeable
effect.  

\subsection{Conservation laws}
\begin{figure}[ht!]
\epsfig{file=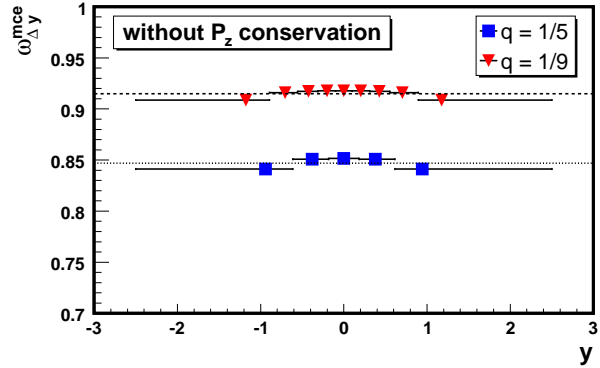,width=8cm}
\caption{Same as Fig.\ref{dodp} right, however without $P_z$
  conservation.} 
\label{dodywopz}
\end{figure}

It seems worthwhile to consider individual conservation laws and their impact
on multiplicity fluctuations in more detail. On of the main advantages of the
analytical procedure presented here, is certainly that one can easily `switch
on` or `switch off` a particular conservation law.  For illustrative purposes
we show the result of $\omega^{mce}_{\Delta y}$ for MCE without longitudinal
momentum conservation in Fig.\ref{dodywopz}.

In comparing Fig.\ref{dodp}, right, to  Fig.\ref{dodywopz} it becomes
obvious that energy conservation alone cannot account for the strong
suppression of multiplicity fluctuations at forward rapidities, but has to be
explained by combined energy and longitudinal momentum conservation. 

The relevant cumulants elements, which give information about the strenght of
correlations between particle number and a particular conserved quantity, are
$\kappa_2^{N_{\Omega},Q}$, $\kappa_2^{N_{\Omega},E}$,
$\kappa_2^{N_{\Omega},p_x}$, etc. Whenever a element is vanishig, then the
corresponding conservation law has no impact on multiplicity fluctuations. For
 details of the calculations please see the Appendix. Since for fluctuations
 of charged particles $\kappa_2^{N_{\Omega},Q}$ and $\kappa_2^{N_{\Omega},E}$
 are generally non-zero, we will focus only on the effects of momentum
 conservation. 

For multiplicity fluctuations in bins in transverse momentum momentum
conservation does not affect the result, see Appendix \ref{App_pt}, and the
suppression effect is a result of energy conservation alone. When considering
cuts in rapidity one finds in general $\kappa_2^{N_y,p_z} \not= 0$, but
$\kappa_2^{N_y,p_x} = \kappa_2^{N_y,p_y} = 0$, and only longitudinal momentum
conservation needs to be taken into account, see Appendix \ref{App_y}. In
considering the third idealized case, where our detector observes only a
segment in azimuthal angle $\phi$, but all rapidities $y$ and transverse
momenta $p_T$, both global $P_x$, and $P_y$ conservation lead to non-trivial
modifications of Eq.(\ref{accscaling}), 
see Appendix \ref{App_phi}. 

To understand the difference between the strong suppression of fluctuations at
high transverse momentum and the rather modest suppression at high rapidity
when momentum conservation is not enforced, one should compare the elements
$\kappa_2^{N_{p_T},E}$ in Eq.(\ref{omega_pt}), and $\kappa_2^{N_{y},E}$ in
Eq.(\ref{omega_y}), which measure in Boltzmann approximation the average
energy density carried by particles in a bin $\Delta p_T$ or $\Delta y$, to
the total average energy density $\langle E_-\rangle =  \kappa_2^{N_{4\pi},E}$
carried by $\pi^-$. (All other elements in Eqs.(\ref{omega_pt}) and
(\ref{omega_y}) do not depend on the location of the segment.) In case of
kinematical cuts in $\Delta p_T$ the fraction $\kappa_2^{N_{p_T},E} / \langle
E_- \rangle$ rises from about $5\%$ in the lowest to roughly $20\%$ in the
highest $p_T$-bin. In contrast to that for the central $y$-bin this ratio is
about $10\%$, while the most forward or backward bins it is roughly
$12\%$. However in both cases the bins contain on average $q=1/9\approx 11\%$
of the total average $\pi^-$ yield. The effect of energy conservation is thus
weaker for cuts in rapidity than for cut in transverse momentum, see also
Appendices \ref{App_pt} and \ref{App_y}.  

\subsection{Charged systems}
\begin{figure}[ht!]
\epsfig{file=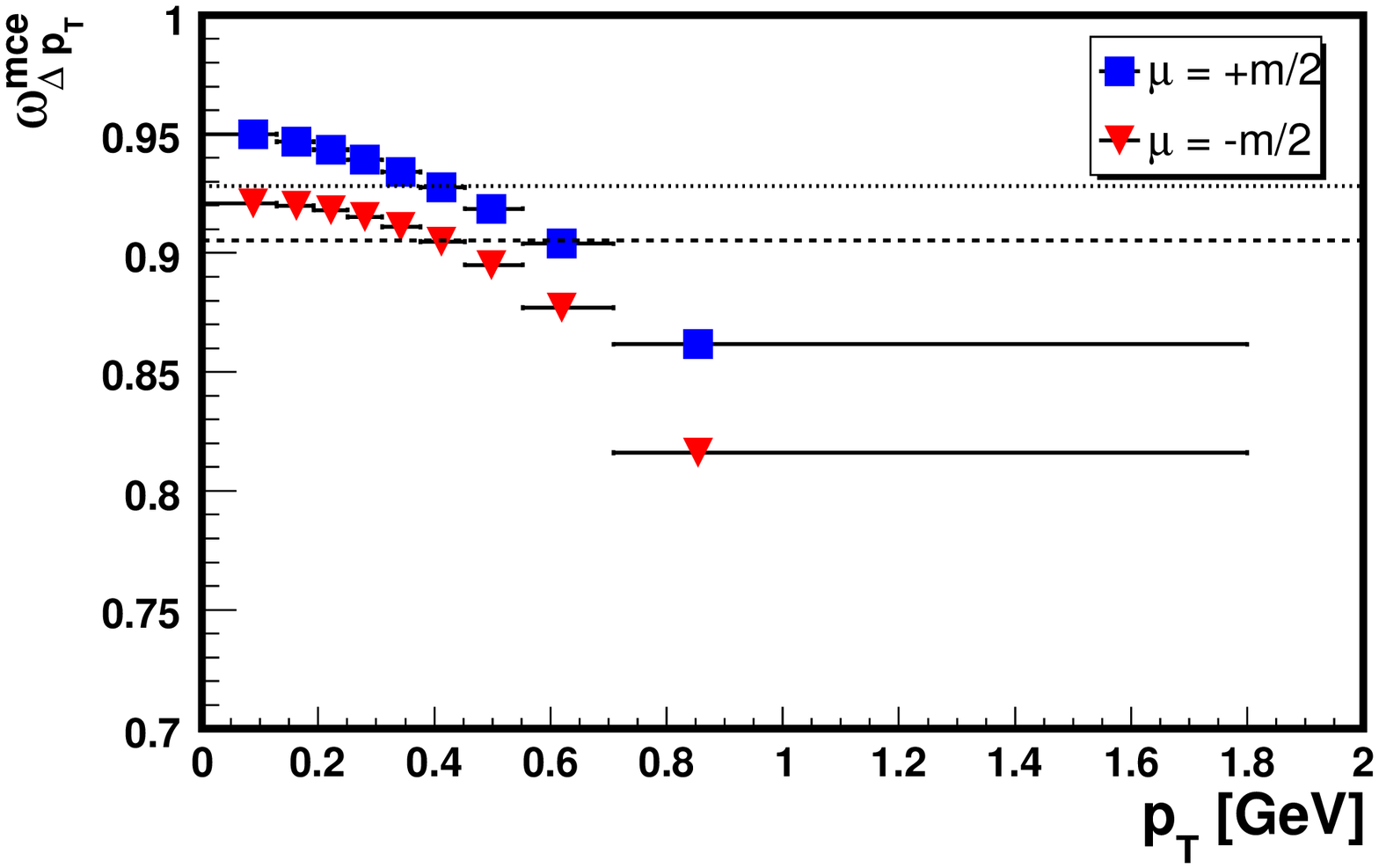,width=8cm}
\epsfig{file=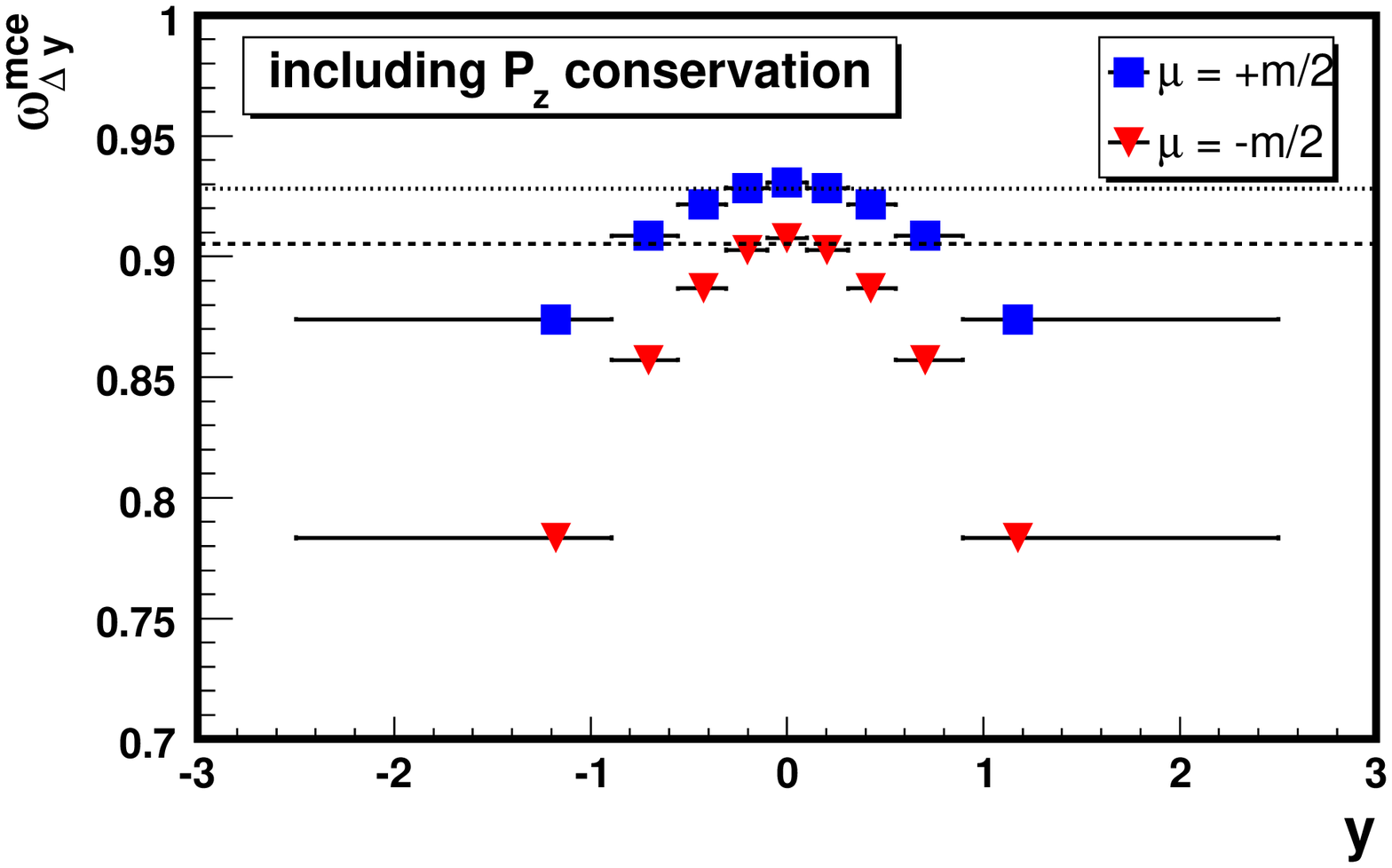,width=8cm}
\caption{ Transverse momentum (left) and rapidity dependence (right) of the
  scaled variance of $\pi^-$ at $T=160MeV$, for a classical pion gas at zero
  charge density. Momentum bins are constructed in a way that each bin
  contains the same fraction $q$ of the average $\pi^-$ yield. The (error)bars
  indicate the width of the $p_T$ or $y$ bins,  while the marker indicate the
  center of gravity of the corresponding bin. The lines indicate acceptance
  scaling  Eq.(\ref{accscaling}). Calculations are done for different charge
  densities $\mu=-\frac{m}{2}$ (square marker, dotted), and $\mu=+\frac{m}{2}$
  (triangle down, dashed).}  
\label{dodp_chr}
\end{figure}

In Figs. \ref{dodp_chr} the transverse momentum (left) and rapidity (right)
dependence of the scaled variance is presented for two different values of
charge density.  
Similar to the CE, in MCE the effective size of the heat and charge bath
matters.  We find that in general MCE effects for negatively charged particles
are stronger (weaker) when the electric charge density is negative
(positive). In the limit of a strongly positively charged system, the $\pi^-$ 
subsystem could be considered as embedded in a large heat, charge, and
momentum bath (provided by $\pi^+$ and $\pi^0$ particles) and MCE effects
would cease. The GCE would here be the appropriate limit. In the opposite
limit of a strongly negatively charged system, charge conservation essentially
becomes equivalent to particle number conservation. This scenario might be
more familiar from textbooks, where the CE is usually understood as the
ensemble with fixed particle number. However here also the same arguments as
above apply, except the effect would be much stronger, and
$\omega^{mce}_{4\pi} = 0$.  

In general one would expect that suppression effects in bins of high
transverse momentum or high values of rapidity are stronger the more
abundant the analyzed particle species is. In the context of heavy ion
collision this implies that MCE effects should be stronger for positively
charged particles than for negatively charged particles, due to the fact that
the created system carries positive net-charge.

Previous work suggests that the asymptotic values for the scaled variance are
indeed reached rather quickly \cite{VolDep} and above results are certainly
applicable to large systems expected to be created in relativistic heavy ion
collisions. 

\section{Remarks and Conclusion}
\label{Remarks}
Some concluding remarks seems to be in order. Although it might seem
inappropriate to use the MCE formulation of a hadron resonance gas model for
calculation of multiplicity fluctuations in heavy ion collisions, as energy
and volume cannot be assumed to be the same in all events, it should be
stressed that GCE and CE still imply a very particular type  of heat (and
momentum) bath, namely an infinite (and ideal) one. This assumption seems to
us even less appropriate. Also the MCE is often understood as the ensemble
with energy (and charge), however not momentum conservation. It is usually
assumed that taking momentum conservation into account  will not affect
fluctuations in the large volume limit. We have shown \cite{clt} in a recent
paper that this is indeed the case, when one assumes information about all
produced particles. However for calculations of multiplicity fluctuations in
arbitrary finite subsystems in momentum space all kinematic conservation laws
need to be taken into account. 
  
In a realistic heavy ion experiment it seems impossible to measure the
entire final state of each collision. The observed subsystem could therefore
be seen as effectively embedded into a (possibly much larger) heat, charge, and
momentum bath. Sometimes it is therefore argued that, when investigating only a
small part of a statistical system (canonical or micro-canonical), one can
ignore correlations of the subsystem under investigation with the remaining
system.  This argument is often applied when considering yields and/or
fluctuations in a limited segment of momentum space. More precisely, usually
the GCE is thought to be the appropriate ensemble to model fluctuations of
particle multiplicity or particle ratios found in some mid-rapidity interval
\cite{GCEfluc}. In this work we have argued that this assumption should be
checked carefully. The GCE is only the correct ensemble to choose, if heat and
charge bath are assumed to be infinite, while the observed subsystem remains
finite. 

Based on our previous line of arguments, one would also expect that strong
collective longitudinal and transverse flow would lead to a strong
correlation of macroscopic subsystems. Longitudinal momentum conservation
implies that when `observing` in an event a final state with a certain small
(large) number of produced particles at very forward rapidity, a similarly
small (large) number of particles should exist at backward
rapidities. Particles in these bins carry substantial longitudinal momenta,
and hence energy. Modest fluctuations in their numbers should therefore induce
stronger fluctuations in the central rapidity region. The same line of
arguments is applicable to the transverse momentum dependence. One would
therefore expect a similar momentum space dependence of experimentally
measured charged particle multiplicity fluctuations as shown in
Figs. \ref{dodp}.   

This argument is additionally supported by UrQMD simulations \cite{UrQMDfluc}.
In transport calculations the produced systems stay far away from global or
local equilibrium \cite{TransportEq} and other (dynamical) mechanisms might
lead to similar effects. On the other hand could one also infer from
\cite{UrQMDfluc} that even in non-equilibrium systems correlations due to
exactly enforced conservation laws determine the general trend, although
transport simulations show, for instance, a very different dependence 
of multiplicity fluctuations on beam energy  \cite{HSDfluc1,HSDfluc2} than
statistical equilibrium models. This should be subject of further
investigation.    

Finally, and most importantly, we want to stress that recently presented
preliminary NA49 analysis of multiplicity fluctuations in certain rapidity
and transverse momentum windows \cite{BeniCoolData} shows qualitatively the
very same trends as they are suggested by the MCE formulation of the
statistical model. Data, UrQMD simulations, and the statistical model exhibit
suppressed multiplicity fluctuations when bins with high transverse momentum
(or high values of rapidity) are compared to bins of same mean multiplicity at
lower transverse momentum (or lower values of rapidity). We are certainly
tempted to interpret this rather unexpected common behavior as a manifestation
of energy and momentum conservation effects.

\section{Summary}   
\label{Summary}
We have discussed the effect of momentum space cuts on multiplicity
fluctuations in the framework of an ideal classical pion gas in the three
standard ensembles, GCE, CE, and MCE. Only in the MCE we expect a momentum
space dependence of multiplicity fluctuations, when comparing intervals of same
average multiplicity. We have shown that even in the thermodynamic limit
energy-momentum conservation can leave a sizable effect in the fluctuation
pattern. 

In a previous publication we have argued that despite the fact one may expect
event-by-event fluctuations of the thermal energy, i.e. the part of the total
energy which goes into thermal particle production rather than collective
expansion, these event-by-event fluctuations remain small compared to energy
fluctuations one would expect from grand canonical and canonical ensembles. In
this work we have shown that energy and momentum conservation lead to a
non-trivial momentum space dependence of the fluctuation pattern. This
argument seems to be strongly supported by data. 

Above results become all the more interesting when compared to models which
seek to describe effects beyond our considerations. In fact our calculations
suggest a  similar strength of respective suppression or enhancement as they
were predicted as signals for the critical point of strongly interacting
matter, the onset of deconfinement, or generally a possible phase
transition. One might also be tempted to argue, that enhanced fluctuations
around mid-rapidity, when compared to a more forward rapidity slice, should be
interpreted as a signal of a phase transition from a quark gluon plasma to a
hadron gas phase, expected to be first realized in the presumably hotter and
denser central rapidity region. However in this case there should be a
non-monotonic variation as center of mass energy of colliding nuclei is
changed. This seems not to be supported by preliminary NA49 data. 

In summary, above results should be treated as a prediction for general
trends of multiplicity fluctuations in limited segments of momentum space. The
existence of this general behavior should be further tested by current
experiments. Observation of effects similar to those of Figs. \ref{dodp} in 
experimental data would, in our opinion, strongly speak in favor of our
hypothesis that fluctuations of extensive observables are indeed dominated by
material and motional conservation laws.   

\begin{acknowledgments} 
We would like to thank  F.~Becattini, V.V.~Begun, M.~Bleicher,
E.L.~Bratkovskaya, W.~Broniowski, L.~Ferroni, M.I.~Gorenstein, M.~Ga\'zdzicki,
S.~H\"aussler, V.P.~Konchakovski, B.~Lungwitz, and G.~Torrieri for fruitful
discussions.    
\end{acknowledgments} 

\appendix
\section{Globally Conserved Quantities}
\label{Calc}
Turning now to calculations of cumulants, Eq.(\ref{kappa_n}), we employ
always coordinates most suitable to our problem. The invariant phase space
element is given by:  
\begin{equation}
\varepsilon ~\frac{dN}{d^3p} ~=~ \frac{dN}{m_T~ dm_T ~dy ~d \phi} ~=~
\frac{dN}{p_T ~dp_T ~dy ~d \phi} ~=~ \varepsilon ~\frac{g}{\left( 2\pi\right)^3}
~\exp \left( -\frac{\varepsilon-\mu}{T}\right)~,
\end{equation}
where the single particle energy $\varepsilon = m_T \cosh y$, its longitudinal
momentum $p_z = m_T \sinh y$, transverse mass $m_T^2 = p_T^2 + m^2 $,
transverse momentum $p_T^2 = p_x^2 + p_y^2$, and rapidity $y = \tanh \left(
  p_z/\varepsilon\right)$. Additionally we employ spherical coordinates:
\begin{equation}
\frac{dN}{d^3p} ~=~ \sin \theta ~ p^2 ~ \frac{dN}{d\phi~ d\theta ~dp}~.
\end{equation}
For clarity we consider explicitely a few terms, not
given in \cite{clt}, here. The total energy density is given by the sum over
individual contributions of all particle species $k$:   
\begin{eqnarray}\label{kappa_1_E}
\kappa_1^E &=& \left(- i \frac{\partial }{\partial
    \phi_{E}}\right) \Psi\left( \phi_j \right) \Bigg|_{\phi_j=\vec 0} ~=
\sum_k \int \limits_{0}^{+ \pi} ~d \theta \int 
\limits_{-\pi}^{+\pi} d \phi \int \limits_{0}^{\infty} ~dp
~\varepsilon_k~  \frac{dN_k}{d\phi ~d\theta ~ dp} \nonumber \\
 &=& \sum_k \frac{g_k ~e^{\frac{q_k\mu}{T}}}{2\pi^2} ~m_k^3 ~T~ \left[K_1
   \left(\frac{m_k}{T} \right) + 3~ \frac{T}{m_k}~ K_2 \left(\frac{m_k}{T}
   \right)\right]~= \sum_k \langle E_k \rangle~.
\end{eqnarray}
The diagonal energy element $\kappa_2^{E,E}$ is given by:
\begin{eqnarray}
\kappa_2^{E,E} &= & \left(- i \frac{\partial }{\partial
    \phi_{E}}\right)^2 \Psi\left( \phi_j \right) \Bigg|_{\phi_j=\vec 0} ~=
\sum_k \int \limits_{0}^{+ \pi} ~d \theta \int 
\limits_{-\pi}^{+\pi} d \phi \int \limits_{0}^{\infty}~dp
~\varepsilon_k^2~  \frac{dN_k}{d\phi ~d\theta ~ dp}\nonumber \\
&=& \sum_k \frac{g_k ~e^{\frac{q_k\mu}{T}}}{2\pi^2} ~m_k^4 ~T~ \left[K_0
  \left(\frac{m_k}{T} \right) + 5~ \frac{T}{m_k} ~K_1 \left(\frac{m_k}{T}
  \right) +  12~\frac{T^2}{m_k^2}~ K_2 \left(\frac{m_k}{T} \right) \right] ~.
\end{eqnarray}
Additionally we define the diagonal momentum correlation terms, with $p_z = p
\cos \theta$: 
\begin{eqnarray}\label{kappa_p_p}
\kappa_2^{p_z,p_z} &=& \left(- i \frac{\partial }{\partial
    \phi_{p_z}}\right)^2 ~\Psi\left( \phi_j \right) \Bigg|_{\phi_j=\vec 0}~=
\sum_k  \int \limits_{0}^{2\pi} d\phi ~ \int \limits_{0}^{\pi} d\theta ~
\int \limits_{0}^{\infty}  dp ~p_z^2 ~\frac{dN_k}{d\phi ~d\theta ~ dp}
\nonumber \\  
&=&\sum_k \frac{g_k ~e^{\frac{q_k\mu}{T}}}{2\pi^2}~ m_k^4~T~
\Bigg[\frac{T}{m_k} ~K_1\left(\frac{m_k}{T} \right) +4~\frac{T^2}{m_k^2}~
K_2\left(\frac{m_k}{T} \right) \Bigg]~.
\end{eqnarray}
Due to spherical symmetry in momentum space we find $\kappa_2^{p_x,p_x} =
\kappa_2^{p_y,p_y} = \kappa_2^{p_z,p_z}$. Correlation terms of odd order in
one of the momenta are identical to zero. As an example we find for
correlations between energy and longitudinal momentum: 
\begin{eqnarray}\label{kappa_E_p}
\kappa_2^{E,p_z} &=& \left(- i \frac{\partial }{\partial
    \phi_{E}}\right) \left(- i \frac{\partial } {\partial
    \phi_{p_z}}\right) \Psi\left( \phi_j \right) \Bigg|_{\phi_j=\vec 0}
\nonumber \\
&=& \sum_k \int \limits_{0}^{2\pi} d\phi  \int \limits_{0}^{\pi} 
d\theta  \int \limits_{0}^{\infty} dp ~\varepsilon ~p_z  ~\frac{dN_k}{d\phi
  ~d\theta ~ dp} =0, 
\end{eqnarray} 
since the integral over the polar angle $\int_{0}^{\pi} \sin \theta \cos
\theta=0$. Similarly we find $\kappa_2^{Q,p_x} = \kappa_2^{p_x,p_y} = 0$. 
Additionally $\kappa_1^{p_x}=0$, etc., since for a static source $\langle \vec
P \rangle = \vec 0$. 

\section{Transverse Momentum Segment}
\label{App_pt}

The average particle number density of $\pi^-$ in a segment of transverse
momentum ${\Delta p_T}$ is given by Eq.(\ref{kappa_n}), i.e. the first
derivative of the CGF with respect to $\phi_{N_{\Omega}}=\phi_{N_{p_T}}$ at
the origin: 
\begin{eqnarray}\label{dNdpt}
\kappa_1^{N_{p_T}} &=&  \left( -i \frac {\partial}
 {\partial \phi_{N_{p_T}}} \right) \Psi\left( \phi_j \right)
\Bigg|_{\phi_j=\vec  0} ~=~ 
\int \limits_{\Delta p_T} dp_T  \int 
\limits_{0}^{2\pi} d\phi  \int \limits_{-\infty}^{\infty} 
dy ~ \frac{dN}{dp_T ~dy ~ d\phi} \nonumber \\
&=& \frac{g~e^{-\frac{\mu}{T}}}{2\pi^2}  \int
\limits_{\Delta p_T} dp_T  ~p_T \sqrt{p_T^2+m^2} ~K_1 \left(
  \frac{\sqrt{p_T^2+m^2}}{T}\right)~.
\end{eqnarray}
Please note that $\kappa_1^{N_{p_T}} = \int_{\Delta p_T} dp_T ~dN/dp_T =
\langle N_{p_T} \rangle$. Correlations of $\pi^-$ in a segment ${\Delta
  p_T}$ with globally conserved energy are given by double differentiation of
$\Psi\left( \phi_j \right)$ with respect to $\phi_{N_{p_T}}$ and $\phi_E$, thus:
\begin{eqnarray}
\kappa_2^{N_{p_T},E} &=&  \left(- i \frac{\partial } {\partial
    \phi_{N_{p_T}}}\right) \left(- i \frac{\partial } {\partial
    \phi_{E}}\right) \Psi\left( \phi_j \right) \Bigg|_{\phi_j=\vec 0}~=  ~ \int
\limits_{\Delta  p_T} dp_T ~ \int \limits_{0}^{2\pi} d\phi ~ \int
\limits_{-\infty}^{\infty} dy ~ \varepsilon ~ \frac{dN}{dp_T ~dy ~ d\phi}
\nonumber \\  
&=& \frac{g~e^{-\frac{\mu}{T}}}{2\pi^2} \! \!\int \limits_{\Delta p_T} dp_T ~
p_T \left(p_T^2+m^2 \right) \left[ K_0 \left( \frac{\sqrt{p_T^2+m^2}}{T}
  \right) + \frac{T}{\sqrt{p_T^2+m^2}} ~ K_1 \left( \frac{\sqrt{p_T^2+m^2}}{T}
  \right) \right] ~.
\end{eqnarray}
Correlations between conserved momenta and particles in $\Delta p_T$, given by
the elements $\kappa_2^{N_{p_T},p_x}$,$\kappa_2^{N_{p_T},p_y}$, and
$\kappa_2^{N_{p_T},p_z} $ are identical to zero, due to symmetry
in azimuthal angle $\phi$ for the first two, and due to an antisymmetric
rapidity integral for the last. Therefore, all elements involving an odd order
in one of the momenta in Eq.(\ref{matrix}) are equal to zero. The determinant
of Eq.(\ref{matrix}) thus factorizes, similar to Eq.(\ref{normdet}), into a
product of $(\kappa_2^{p_x,p_x})^3$ and the determinant of a $3
\times 3$ sub-matrix involving only terms containing $N_{p_T},E,Q$. Hence
momentum conservation drops out when calculating
Eq.(\ref{SimpleOmega}). However the strength of correlations between particle
number $N_{p_T}$ and globally conserved energy $E$ will depend on the position
of the segment $\Delta p_T$. Thus using Eqs.(\ref{SimpleOmega}) and
(\ref{matrix}), one can express the width of the MCE multiplicity distribution
(\ref{PMCE}) by: 
\begin{equation}\label{omega_pt}
  \omega^{mce}_{\Delta p_T}~=~
  \frac{\kappa_2^{N_{p_T},N_{p_T}}}{\kappa_1^{N_{p_T}}} - 
  \frac{1}{\kappa_1^{N_{p_T}} \det |\hat \kappa_2|}\Bigg[ 
  \left( \kappa_2^{N_{p_T},Q} \right)^2 \kappa_2^{E,E}+
  \left( \kappa_2^{N_{p_T},E} \right)^2 \kappa_2^{Q,Q}  - 2
  \kappa_2^{N_{p_T},E} \kappa_2^{N_{p_T},Q} \kappa_2^{E,Q} \Bigg]  
\end{equation} 
In Boltzmann approximation, we find from Eq.(\ref{kappa_n}),
$\kappa_2^{N_{p_T},N_{p_T}} = \kappa_2^{N_{p_T},Q} = \kappa_1^{N_{p_T}} = q
\kappa_1^{N_{4\pi}}$, where we have defined the acceptance
$q \equiv \kappa_1^{N_{p_T}} / \kappa_1^{N_{4\pi}} $. However, when observing a
fraction $q$ of the particle density, one does not necessarily 
observe the same fraction $q$ of the energy density $\langle E_- \rangle$
carried by $\pi^-$, and thus $\kappa_2^{N_{p_T},E} \not= q \langle E_-
\rangle$. Therefore depending on the location of $\Delta p_T$, our detector
sees a larger (smaller) fraction of the total energy, which leads to smaller
(larger) particle number fluctuations, see Fig. \ref{dodp}, left panel. One
can easily verify that setting $\kappa_2^{N_{p_T},E} = q \langle E_- \rangle $ in
Eq.(\ref{omega_pt}), leads to acceptance scaling, Eq.(\ref{accscaling}),
$\omega^{mce}_{\Delta p_T} = 1 + q \left(\omega^{mce}_{4\pi}-1 \right)$.  

\section{Rapidity Segment}
\label{App_y}

The average particle number density of $\pi^-$ in a rapidity interval
${\Delta y}$ is given by: 
\begin{eqnarray}\label{dNdy}
\kappa_1^{N_y} &=&  \left(- i \frac{\partial } {\partial
    \phi_{N_{y}}}\right) \Psi\left( \phi_j \right) \Bigg|_{\phi_j=\vec 0}~=
~\int \limits_{m}^{\infty} dm_T ~ \int \limits_{0}^{2\pi} d\phi ~ \int
\limits_{\Delta y} dy  ~ \frac{dN}{dm_T ~dy ~ d\phi}  \nonumber \\
&=& \frac{g~e^{-\frac{\mu}{T}}}{\left(2\pi \right)^2}~ T^3~\int
\limits_{\Delta y} dy~ \exp \left( -\frac{m }{T} \cosh \left(y \right) \right)
\left[ \left(\frac{m}{T}\right)^2 + 2\frac{m}{T} \cosh^{-1} y  +
  2 \cosh^{-2} y  \right]~.
\end{eqnarray}
Please note that $\kappa_1^{N_{y}} = \int_{\Delta y} dy ~dN/dy =
\langle N_{y} \rangle$. Correlations of particles in ${\Delta y}$ with
globally conserved energy are given by:
\begin{eqnarray}
\kappa_2^{N_y,E} &=& \left(- i \frac{\partial } {\partial
    \phi_{N_{y}}}\right) \left(- i \frac{\partial } {\partial
    \phi_{E}}\right) \Psi\left( \phi_j \right) \Bigg|_{\phi_j=\vec 0}~=  ~\int
\limits_{m}^{\infty} dm_T ~ \int \limits_{0}^{2\pi} d\phi ~ \int
\limits_{\Delta y} dy ~ \varepsilon ~ \frac{dN}{dm_T ~dy ~ d\phi} \nonumber \\ 
&=&\frac{g~e^{-\frac{\mu}{T}}}{\left(2\pi \right)^2} ~ T^4 ~ \int
\limits_{\Delta y} dy ~ \cosh y ~\exp \left(- \frac{m}{T} \cosh  y \right) 
\nonumber \\ && \times~ \left[\left(\frac{m}{T} \right)^3  + 3 \left(
    \frac{m}{T} \right)^2 \cosh^{-1} y+ 6 ~\frac{m}{T}~ \cosh^{-2} y +   6
  \cosh^{-3} y \right]~.
\end{eqnarray}
The correlation term of particles in ${\Delta y}$ with globally conserved
longitudinal momentum $P_z$ reads:
\begin{eqnarray}
\kappa_2^{N_y,p_z} &=& \left(- i \frac{\partial } {\partial
    \phi_{N_{y}}}\right) \left(- i \frac{\partial } {\partial
    \phi_{p_z}}\right) \Psi\left( \phi_j \right) \Bigg|_{\phi_j=\vec 0}~= \int
\limits_{0}^{2\pi} d\phi ~ \int \limits_{m}^{\infty} dm_T ~  \int 
\limits_{\Delta y} dy ~ p_z ~\frac{dN}{dm_T ~dy ~ d\phi} \nonumber \\ 
&=& \frac{g~e^{-\frac{\mu}{T}}}{\left(2\pi \right)^2} ~ T^4 ~ \int
\limits_{\Delta y} dy ~ \sinh y ~ \exp \left(- \frac{m}{T} \cosh y \right)   
\nonumber \\ &&\times~
\left[\left(\frac{m}{T} \right)^3 + 3 \left( \frac{m}{T} \right)^2
  \cosh^{-1}   y + 6 ~\frac{m}{T}~ \cosh^{-2} y +   6 \cosh^{-3} y \right]~.
\end{eqnarray}
Thus the element $\kappa_2^{N_y,p_z}$ in the matrix (\ref{matrix}) is
non-vanishing, and longitudinal momentum ($P_z$) conservation seems to affects
correlations between particles in a segment $\Delta y$ and the remaining
system. In contrast to that further elements are equal to zero,
$\kappa_2^{N_y,p_x} =\kappa_2^{N_y,p_y} = 0$, and $P_x$ and $P_y$ conservation
have no additional effect. When momentum conservation is taken into account
the scaled variance (\ref{SimpleOmega}) can be calculated from
Eq.(\ref{calcdet}):  
\begin{eqnarray}\label{omega_y}
  \omega^{mce}_{\Delta y} = \frac{\kappa_2^{N_{y},N_{y}}}{\kappa_1^{N_{y}}}
  &-& \frac{1}{\kappa_1^{N_{y}} \kappa_2^{p_z,p_z} \det |\hat \kappa_2|}\Bigg[ 
  \left( \kappa_2^{N_{y},Q} \right)^2 \kappa_2^{E,E} \kappa_2^{p_z,p_z}+
  \left( \kappa_2^{N_{y},E} \right)^2 \kappa_2^{Q,Q} \kappa_2^{p_z,p_z}
  \nonumber \\
  &+&\left( \kappa_2^{N_{y},p_z} \right)^2 \left[ \kappa_2^{Q,Q}\kappa_2^{E,E}
    -  \left( \kappa_2^{E,Q} \right)^2 \right] -
  2 \kappa_2^{p_z,p_z}  \kappa_2^{N_{y},E} \kappa_2^{N_{y},Q}
  \kappa_2^{E,Q}~\Bigg]~. 
\end{eqnarray} 
Similarly to the previous section, we find a large (small)
$\kappa_2^{N_{y},p_z}$  leads to small (large) fluctuations, see
Fig. \ref{dodp}, right panel. When intervals symmetric in rapidity are
assumed, e.g. $\Delta y = \left[-y_1,y_1 \right]$, or $ \Delta y =
\left[-y_2,-y_1\right] \cup \left[y_1,y_2 \right]$, correlations between
particle number and momentum disappear, $\kappa_2^{N_{y},p_z}=0$, and
Eq.(\ref{omega_y}) reduces to Eq.(\ref{omega_pt}), and momentum conservation
does not play a role. Equally when disregarding longitudinal momentum
conservation the same arguments as those of Appendix \ref{App_pt} apply and
Eq.(\ref{omega_pt}) holds, however the effect is much weaker, see
Fig. \ref{dodywopz}. 
 
\section{Azimuthal Angle Segment}
\label{App_phi}
Th average particle number in $\Delta \phi $, while integrating over all $p_T$
and $y$ is simply a fraction $q = \Delta \phi / 2 \pi$ of the total
yield $\langle N_{4\pi}\rangle$. Therefore  $\kappa_1^{N_{\phi}} = q
\kappa_1^{N_{4\pi}}$. Equally, the energy carried by $\pi^-$ in this
interval is $\kappa_2^{E,N_{\phi}}= q \langle E_-\rangle$. Due to symmetry
around $y=0$, we find additionally $\kappa_2^{N_{\phi},p_z}= 0$. However for
the transverse momenta $p_x= p_T \cos \phi$, and $p_y= p_T \sin \phi$ the
correlation with $N_{\phi}$ is generally non-zero. 
\begin{eqnarray}
\kappa_2^{N_{\phi},p_x} &=& \left(- i \frac{\partial } {\partial
    \phi_{N_{\phi}}}\right) \left(- i \frac{\partial } {\partial
    \phi_{p_x}}\right) \Psi\left( \phi_j \right) \Bigg|_{\phi_j=\vec
  0}~=\frac{g}{\left(2\pi \right)^3} \int 
\limits_{\Delta \phi} d\phi ~ \int \limits_{0}^{\infty} dp_T ~  \int
\limits_{-\infty}^{\infty} dy ~ p_x ~\frac{dN}{dp_T ~dy ~ d\phi} 
\nonumber \\ &=& 
\int \limits_{\Delta \phi}d\phi \cos \phi ~
~\frac{2g~e^{-\frac{\mu}{T}}}{\left(2\pi 
  \right)^3}~m^2~T~\sqrt{\frac{\pi}{2}~m~T} ~ K_{5/2} 
\left( \frac{m}{T} \right)~=~ \left( 2\pi
\right)^{-1} \Big[ \sin \phi \Big]_{\Delta \phi}  ~\langle p_T \rangle ~.
\end{eqnarray}
Similarly we find $\kappa_2^{N_{\phi},p_y} = - \left( 2\pi
\right)^{-1} \left[ \cos \phi \right]_{\Delta
  \phi}  ~\langle p_T \rangle $. Unlike in the previous sections there is no 
particular dependence of the position of the interval $\Delta \phi$. However
in general there is a dependence. When momentum conservation is taken into
account Eq.(\ref{SimpleOmega}) can be calculated from Eq.(\ref{calcdet}):
\begin{eqnarray}\label{omega_phi}
  \omega^{mce}_{\Delta \phi} &=&
  \frac{\kappa_2^{N_{\phi},N_{\phi}}}{\kappa_1^{N_{\phi}}} - 
  \frac{1}{\kappa_1^{N_{\phi}} \kappa_2^{p_x,p_x} \det |\hat
    \kappa_2|} \Bigg[ \left( \kappa_2^{N_{\phi},Q} \right)^2 \kappa_2^{E,E} 
    \kappa_2^{p_x,p_x}  +  \left( \kappa_2^{N_{\phi},E} \right)^2
    \kappa_2^{Q,Q} \kappa_2^{p_x,p_x} 
  \nonumber \\ 
  &+&\left( \left( \kappa_2^{N_{\phi},p_x} \right)^2 +  \left(
      \kappa_2^{N_{\phi},p_y} \right)^2  \right)  \left[
    \kappa_2^{Q,Q}\kappa_2^{E,E} 
    -  \left( \kappa_2^{E,Q} \right)^2 \right] -
  2  \kappa_2^{p_x,p_x}  \kappa_2^{N_{\phi},E} \kappa_2^{N_{\phi},Q}
  \kappa_2^{E,Q}~\Bigg]~, 
\end{eqnarray} 
where we have used $\kappa_2^{p_x,p_x}=\kappa_2^{p_y,p_y}$.
As mentioned before there is no particular dependence of
$\kappa_2^{N_{\phi},E}$ and $\kappa_2^{N_{\phi},Q}$ on the position of $\Delta
\phi$. However we have a term $ \left( \kappa_2^{N_{\phi},p_x} \right)^2 +
\left( \kappa_2^{N_{\phi},p_y} \right)^2 $. In case we assume a continuous
interval $\Delta \phi_A = \left[\phi_1,\phi_2  \right]$ this terms reads:
\begin{equation}
\left( \kappa_2^{N_{\phi},p_x} \right)^2 +  \left( \kappa_2^{N_{\phi},p_y}
\right)^2  =  \frac{\langle p_T \rangle^2}{\left( 2\pi \right)^2} ~\left[
 1- \cos \left( \phi_1 - \phi_2 \right) \right]
\end{equation}
This term is evidently positive, hence fluctuations are suppressed. One can
easily verify that when one takes $\Delta \phi_B =\left[\phi_1,\phi_2\right]
\cup \left[\phi_1+\pi,\phi_2+\pi \right]$, i.e. two opposite slices in
azimuthal angle, the correlation disappears,  $\kappa_2^{N_{\phi},p_x}
=\kappa_2^{N_{\phi},p_y}= 0 $, and one returns to acceptance scaling,
Eq.(\ref{accscaling}).   


\end{document}